\documentclass[11pt]{article}
\usepackage[margin=1in]{geometry}

\usepackage{times,epsfig}
\usepackage{graphicx}
\usepackage{subfig}
\usepackage{stfloats}
\usepackage{algorithmic}
\usepackage{algorithm}

\usepackage{amsmath}
\usepackage{amssymb}
\usepackage{amsfonts}

\usepackage{verbatim}
\usepackage{appendix}
\usepackage{amsthm}
\usepackage{url}
\usepackage{epstopdf}
\usepackage{hyperref}
\usepackage{array}

\usepackage{ifthen}
\newboolean{includeall}
\setboolean{includeall}{false}

\newboolean{fullpaper}
\setboolean{fullpaper}{true}

\newcommand{\condcomment}[3]{\ifthenelse{#1}{#2}{#3}}

\DeclareCaptionType{copyrightbox}

\newtheorem{ndef}{Definition}
\newtheorem{nlem}{Lemma}
\newtheorem{nexp}{Example}
\newtheorem{nprop}{Proposition}
\newtheorem{nthm}{Theorem}

\newcommand{\nop}[1]{}

\newenvironment{proofsketch}{\noindent{\em Proof Sketch.~}}{\qed\vspace{1em}} 

\begin{document}
\title{On Load Shedding in Complex Event Processing}

\author{Yeye He\footnote{Microsoft Research, Redmond, WA. {\tt yeyehe@microsoft.com}} \and Siddharth Barman\footnote{California Institute of Technology, Pasadena, CA. {\tt barman@caltech.edu}} \and Jeffrey F. Naughton\footnote{University of Wisconsin--Madison, Madison, WI. {\tt naughton@cs.wisc.edu}}}

\date{}


\maketitle

\begin{abstract}
Complex Event Processing (CEP) is a stream processing model that
focuses on detecting event patterns in continuous event streams.
While the CEP model has gained popularity in the research communities and
commercial technologies, the problem of gracefully degrading
performance under heavy load in the presence of resource constraints,
or load shedding, has been largely overlooked.  CEP is similar to
``classical'' stream data management, but addresses a substantially
different class of queries. This unfortunately renders the load
shedding algorithms developed for stream data processing
inapplicable.  In this paper we study CEP load shedding under
various resource constraints.  We formalize broad classes of CEP 
load-shedding scenarios as different optimization problems. We demonstrate
an array of complexity results that reveal the hardness of
these problems and construct shedding algorithms with performance
guarantees. Our results shed some light on the difficulty of
developing load-shedding algorithms that maximize utility.
\end{abstract}

\vspace{-0.2cm}
\section{Introduction}
\label{sec:introduction}

The complex event processing or CEP model has received significant
attention from the research community~\cite{Agrawal08, Gyllstrom07,
  Liu09, Mei09, White07, Wu06}, and has been adopted by a number of
commercial systems including Microsoft
StreamInsight~\cite{streaminsight}, Sybase Aleri~\cite{sybase}, and
StreamBase~\cite{streambase}.  A wide range of applications, including
inventory management~\cite{wavemark}, behavior
monitoring~\cite{wang11}, financial trading~\cite{Adi12}, and fraud
detection~\cite{Aniello11, Widder09}, are now powered by CEP
technologies.

A reader who is familiar with the extensive stream data processing
literature may wonder if there is anything new here, or if CEP
is just another name for stream data management. While both
kinds of system evaluate queries over data streams, the important
difference is the class of queries upon which each system
focuses. In the traditional stream processing literature, the focus
is almost exclusively on aggregate queries or binary equi-joins. By
contrast, in CEP, the focus is on detecting certain
patterns, which can be viewed as multi-relational non-equi-joins on the
time dimension, possibly with temporal ordering constraints. The class of queries
addressed by CEP systems requires different evaluation algorithms and
different load-shedding algorithms than the class previously considered
in the context of stream data management.

As an example of a CEP-powered system, consider the health care
monitoring system, HyReminder~\cite{wang11}, currently deployed at the
University of Massachusetts Memorial Hospital. The HyReminder system
tracks and monitors the hygiene compliance of health-care workers.  In
this hospital setting, each doctor wears an RFID badge that can be
read by sensors installed throughout the hospital. As doctors walk
around the hospital, the RFID badges they wear trigger ``event'' data,
which is transmitted to a central CEP engine. The CEP engine in 
turn looks for patterns to check for 
hygiene compliance.  As one example, according to the
US Center for Disease Control (CDC)~\cite{jm2002}, a doctor who enters
or exits a patient room (which is captured by sensors installed in the
doorway and encoded as an {\small\texttt{Enter-Patient-Room}} event or
{\small\texttt{Exit-Patient-Room}} event) should cleanse her hands
(encoded by a {\small\texttt{Sanitize}} event) within a short period
of time.  This hygiene regulation can be tracked and enforced using
the following CEP queries. \\ {\small\texttt{Q1: SEQ({\small\texttt{Sanitize}},
    {\small\texttt{Enter-Patient-Room}}) within 1 min}}
\\ {\small\texttt{Q2: SEQ({\small\texttt{Exit-Patient-Room}},
    {\small\texttt{Sanitize}}) within 1 min}} \\ In the HyReminder
system, these two CEP queries monitor the event sequence to track
sanitization behavior and ensure hygiene compliance.
As another example consider CIMS~\cite{wavemark}, which is a system also
powered by CEP and deployed in the same University of Massachusetts
hospital. CIMS is used for inventory management and asset tracking
purposes.  It captures RFID events triggered by tags attached to
medical equipment and expensive medicines, and uses CEP to
track supply usage and reduce inventory cost~\cite{carter09}.

While the emergence of CEP model has spawned a wide variety of
applications, so far research efforts have focused almost exclusively
on improving CEP query join efficiency~\cite{Agrawal08, Gyllstrom07,
  Liu09, Mei09, White07, Wu06}.  \textit{Load shedding}, an important
issue that has been extensively studied in traditional stream
processing~\cite{Babcock03a, Das03, Gedik08, Gedik07, Srivastava04,
  Tatbul03, Wei08, Wei10, Xing06}, has been largely overlooked in the
new context of CEP.

Like other stream systems, CEP systems often face bursty input data.
Since over-provisioning the system to the point where it can handle
any such burst may be uneconomical or impossible, during peak loads a
CEP system may need to ``shed'' portions of the load.  The key
technical challenge herein is to selectively shed work so as to
eliminate the less important query results, thereby preserve 
the more useful query results as defined by some \textit{utility function}.

More specifically, the problem we consider is the following. Consider
a CEP system that has a number of pattern queries, each of which
consists of a number of events and is associated with a
\textit{utility function}. During peak loads, memory and/or CPU
resources may not be sufficient.  A utility-maximizing load shedding
scheme should then determine which events should be preserved in
memory and which query results should be processed by the CPU, so
that not only are resource constraints respected, but also the overall utility
of the query results generated is maximized.

We note that, in addition to utility-maximization in a single CEP
application, the load-shedding framework may also be relevant to cloud
operators that need to optimize across multiple CEP applications.
Specifically, stream processing applications are gradually shifting to
the cloud~\cite{Kleiminger11}.  In cloud scenarios, the cloud operator
in general cannot afford to provision for the aggregate peak load
across all tenants, which would defeat the purpose of
consolidation. Consequently, when the load exceeds capacity, cloud
operators are forced to shed load. They have a financial incentive to
judiciously shed work from queries that are associated with a low
penalty cost as specified in Service Level Agreements (SLAs), so that
their profits can be maximized (similar problems have been called
``profit maximization in a cloud'' and have been considered in the
Database-as-a-Service literature~\cite{chi11a, chi11b}).  Note that
this problem is naturally a utility-maximizing CEP load shedding
problem, where \textit{utility} essentially becomes the financial
rewards and penalties specified in SLAs.

While load shedding has been extensively studied in the context of
general stream processing~\cite{Babcock03a, Das03, Gedik08, Gedik07,
  Srivastava04, Tatbul03, Wei08, Wei10, Xing06}, the focus there is
aggregate queries or two-relation equi-join queries, which are
important for traditional stream joins.  The emerging CEP model,
however, demands multi-relational joins that predominantly use
non-equi-join predicates on timestamp. As we will discuss in more
detail in Section~\ref{sec:related}, the CEP load shedding problem is
significantly different and considerably harder than the problems
previously studied in the context of general stream load shedding.

We show in this work that variants of the utility maximizing load
shedding problems can be abstracted as different optimization
problems.  For example, depending on which resource is constrained, we
can have three problem variants, namely \textit{CPU-bound load
  shedding}, \textit{memory-bound load shedding}, and
\textit{dual-bound load shedding} (with both CPU- and
memory-bound). In addition we can have \textit{integral load
  shedding}, where event instances of each type are either all
preserved in memory or all shedded; and \textit{fractional load
  shedding}, where a sampling operator exists such that a fraction of
event instances of each type can be sampled according to a
predetermined sampling ratio.  Table~\ref{tab:variants} summarizes the
six variants of CEP load shedding studied in this paper: IMLS
(integral memory-bound load shedding), FMLS (fractional memory-bound
load shedding), ICLS (integral CPU-bound load shedding), FCLS
(fractional CPU-bound load shedding), IDLS (integral dual-bound load
shedding), and FDLS (fractional dual-bound load shedding).

\begin{table}
\centering
 \small	\begin{tabular}
{|c|c|c|c|} \hline
 & Memory-bound & CPU-bound  & Dual-bound  \\ \hline %
Integral & IMLS & ICLS  & IDLS \\ \hline %
Fractional &  FMLS & FCLS & FDLS   \\ \hline %
\end{tabular}
\vspace{-0.2cm}
\caption{Problem variants for CEP load shedding} \label{tab:variants}
\vspace{-0.7cm}
\end{table}

We analyze the hardness of these six variants, and study efficient
algorithms with performance guarantees.  We demonstrate an array of
complexity results.  In particular, we show that CPU-bound load
shedding is the easiest to solve: FCLS is solvable in polynomial time,
while ICLS admits a FPTAS approximation.  For memory-bound problems, we
show that IMLS is in general NP-hard and hard to approximate.  We then
identify two special cases in which IMLS can be efficiently
approximated or even solved exactly, and describe a general rounding
algorithm that achieves a
bi-criteria approximation. As for the fractional FMLS, we show it is
hard to solve in general, but has approximable special cases.
Finally, for dual-bound load shedding IDLS and FDLS, we show that they
generalize memory-bound problems, and the hardness results from
memory-bound problems naturally hold. On the positive side, we
describe a tri-criteria approximation algorithm and an approximable
special case for the IDLS problem.

The rest of the paper is organized as follows. We first describe
necessary background of CEP in Section~\ref{sec:background}, and
introduce the load shedding problem in Section~\ref{sec:problem}.  We
describe related work in Section~\ref{sec:related}.  In
Section~\ref{sec:memory}, Section~\ref{sec:cpu} and
Section~\ref{sec:dual} we discuss the memory-bound, CPU-bound and
dual-bound load-shedding problems, respectively. We conclude in
Section~\ref{sec:conclusions}.

\vspace{-0.2cm}
\section{Background: Complex Event Processing}
\label{sec:background}

The CEP model has been proposed and developed by a number of seminal
papers (see~\cite{Agrawal08} and~\cite{Wu06} as examples).  To make
this paper self-contained, we briefly describe the CEP model and its
query language in this section.

\vspace{-0.1cm}
\subsection{The data model}
\vspace{-0.1cm}

Let the domain of possible event types be the alphabet of fixed size $\Sigma =
\{E_i\}$, where $E_i$ represents a type of event. The event stream is
modeled as an event sequence as follows.

\begin{ndef}
\vspace{-0.1cm}
\label{def:sequence}
An \textbf{event sequence} is a sequence $S = (e_{{1}},
e_{{2}},$ $ ... , e_{{N}})$, where each \textbf{event instance} 
$e_{{j}}$ belongs to an \textbf{event type} $E_i \in \Sigma$, 
and has a unique time stamp
$t_{j}$. The sequence $S$ is temporally ordered, that is $t_{j} <
t_{k}$, $\forall j < k$.
\vspace{-0.1cm}
\end{ndef}

\begin{nexp}
\vspace{-0.1cm}
\label{exp:data}

Suppose there are five event types denoted by upper-case characters
$\Sigma $ $= \{A, B, C, D, E\}$. Each character represents a certain
type of event. For example, for the hospital hygiene monitoring
application HyReminder, event type $A$ represents all event instances
where doctors enter ICU, $B$ denotes the type of events where doctors
washing hands at sanitization desks, etc.

A possible event sequence is $S = $ $(A_{1}, B_{2}, $ $C_{3}, D_{4},
E_5, $ $A_{6}, B_{7}, $ $C_{8}, $ $D_9, E_{10})$, where each character
is an instance of the corresponding event type occurring at time stamp
given by the subscript.  So $A_1$ denotes that at time-stamp $1$, a
doctor enters ICU.  $B_2$ shows that at time $2$, the doctor sanitizes
his hands.  At time-stamp 6, there is another enter-ICU event $A_6$,
so on and so forth.
\vspace{-0.2cm}
\end{nexp}

Following the standard practice of the CEP literature~\cite{Agrawal08,
  Ali09, Mei09, Wu06}, we assume events are temporally ordered by
their timestamps. Out-of-order events can be handled using techniques
from~\cite{Chandramouli10, Liu09}.

\begin{ndef}
\vspace{-0.1cm}
\label{def:subsequence}
Given an event sequence $S = (e_{{1}},$ 
$e_{{2}}, ...,$  $e_{{N}})$, a sequence $S' =$ $ (e_{i_{1}}$ $, e_{i_{2}},
... , $ $e_{i_{m}})$ is a \textbf{subsequence} of $S$,
if $1 \leq i_{1} < i_{2} ... < i_{m} \leq N$.
\vspace{-0.1cm}
\end{ndef}

Note that the temporal ordering in the original sequence is preserved
in subsequences, and a subsequence does not have to be
a contiguous subpart of a sequence.

\vspace{-0.1cm}
\subsection{The query model}

Unlike relational databases, where queries are typically ad-hoc and
constructed by users at query time, CEP systems are more like other
stream processing systems, where queries are submitted ahead of time
and run for an extended period of time (thus are also known as
\textit{long-standing queries}).  The fact that CEP queries are known
a priori is a key property that allows queries to be analyzed and
optimized for problems like utility maximizing load shedding.

Denote by $\mathcal{Q} = \{Q_i\}$ the set of CEP queries, where each
query $Q_i$ is a sequence query defined as follows.

\begin{ndef}
\vspace{-0.15cm}
\label{def:query}
A CEP \textbf{sequence query} $Q$ is of the form $Q = SEQ( q^{}_{1},
q^{}_{2}, ... q^{}_{n} )$, where $q^{}_{k} \in \Sigma$ are event
types. Each query $Q$ is associated with a time based sliding window
of size $T(Q) \in \mathbb{R}^{+}$, over which $Q$ will be evaluated.
\vspace{-0.1cm}
\end{ndef}

We then define the \textsf{skip-till-any-match} query match semantics.

\begin{ndef}
\vspace{-0.1cm}
In \textsf{skip-till-any-match}, a subsequence $S' $ $= (e_{i_{1}},
e_{i_{2}}, \ldots, e_{i_{n}})$ of $S$ is considered a \textbf{query match}
of $Q = ( q^{}_{1}, q^{}_{2},$ $\ldots, q^{}_{n} )$ over time window
$T(Q)$ if:

(1) Pattern matches: Event $e_{i_{l}}$ in $S'$ is 
of type $q^{}_{l}$ for all $l \in [1, n]$, 

(2) Within window: $t_{i_{n}} - t_{i_{1}} 
\leq T(Q)$.

\vspace{-0.2cm}
\end{ndef} 

We illustrate query matches using Example~\ref{exp:query}.

\begin{nexp}
\vspace{-0.1cm}
\label{exp:query}

We continue with Example~\ref{exp:data},
where the event sequence
$S = $ $(A_{1}, B_{2}, $ $C_{3},
D_{4}, E_5, $ $A_{6}, B_{7}, $ $C_{8}, $ 
$D_9, E_{10})$.
Suppose there are a total of three queries:
$Q_1$ = $SEQ(A, C)$,
$Q_2$ = $SEQ(C, E)$,
$Q_3$ = $SEQ(A, B, C, D)$,
all having the same window 
size $T(Q_1)$ = $T(Q_2)$ =$T(Q_3) = 5$.

Both sequences $(A_1, C_3)$ and $(A_6, C_8)$ are matches for $Q_1$,
because they match patterns specified in $Q_1$, and are within the
time window 5. However, $(A_1, C_8)$ is not a match even though it
matches the pattern in $Q_1$, because the time difference between
$C_8$ and $A_1$ exceeds the window limit 5.

Similarly, $(C_3, E_5)$ and $(C_8, E_{10})$ are matches of 
$Q_2$; $(A_1,$ $B_2, $ $C_3, D_4)$ and 
$(A_6, B_7, C_8, D_9)$ are matches of $Q_3$.
\vspace{-0.1cm}
\end{nexp}

A query with the \textsf{skip-till-any-match} semantics essentially
looks for the conjunction of occurrences of event types in a
specified order within a time window.  Observe that in
\textsf{skip-till-any-match} a subsequence does not have to be
contiguous in the original sequence to be considered as a match (thus
the word \textit{skip} in its name). Such queries are widely
studied~\cite{streambase, Agrawal08, Gyllstrom07, Liu09,
  Mei09,White07, Wu06} and used in CEP systems.

We note that there are three additional join semantics defined
in~\cite{Agrawal08}, namely, \textsf{skip-till-next-match},
\textsf{partition-contiguity} and \textsf{contiguity}.
\condcomment{\boolean{fullpaper}}
{In the interest of space and also to better focus on the topic of load
shedding, the details of these join semantics are described in
Appendix~\ref{sec:semantics}.}
The load shedding problem formulated in
this work is agnostic of the join semantics used.
\condcomment{\boolean{fullpaper}}
{}
{More details of other join semantics and their relationship with
load shedding appear in the full version of the paper.}

We also observe that although there are CEP language extensions like
negation~\cite{Gyllstrom07} and Kleene closure~\cite{Diao07}, in this
work we only focus on the core language constructs that use
conjunctive positive event occurrences.  We leave such query extensions for
load shedding as future work.

\vspace{-0.2cm}
\section{CEP Load Shedding}
\vspace{-0.1cm}
\label{sec:problem}

It is well known that continuously arriving stream data is often
bursty~\cite{Das03, Srivastava04, Wei10}.  During times of peak loads
not all data items can be processed in a timely manner under resource
constraints. As a result, part of the input load may have to be
discarded (shedded), resulting in the system retaining only a subset
of data items.  The question that naturally arises is which queries
should be preserved while others shedded? To answer this question, we
introduce the notion of utility to quantify the ``usefulness''
of different queries.

\vspace{-0.1cm}
\subsection{A definition of utility}
\label{subsec:utility}
In CEP systems, different queries naturally have different real-world
importance, where some query output may be more important than others.
For example, in the inventory management application~\cite{wavemark,
  carter09}, it is naturally more important to produce real-time query
results that track expensive medicine/equipment than less expensive
ones. Similarly, in the hospital hygiene compliance
application~\cite{wang11}, it is more important to produce real-time
query results reporting serious hygiene violations with grave health
consequences than the ones merely reporting routine compliance.

We define \textit{utility weight} for each query to measure its
importance.

\begin{ndef}
\vspace{-0.15cm}
Let $\mathcal{Q} = \{Q_i\}$ be the set of queries.  Define the
\textbf{utility weight} of query $Q_i$, denoted by $w_i \in
\mathbb{R}^+$, as the perceived usefulness of reporting one instance
of match of $Q_i$.
\vspace{-0.1cm}
\end{ndef}

A user or an administrator familiar with the application can typically
determine utility weights.  Alternatively, in a cloud environment,
operators of multi-tenancy clouds may resort to
service-level-agreements (SLAs) to determine utility weights. In this
work we simply treat utility weights as known constants.  We note that
the notion of query-level weights has been used in other parts of data
management literature (e.g., query scheduling~\cite{Narayanan11}).

The total utility of a system is then defined as follows.

\begin{ndef}
\vspace{-0.15cm}
\label{def:utility}
Let $\mathcal{C}(Q_i, S)$ be the number of distinct matches for query
$Q_i$ in $S$. The \textbf{utility} generated for query $Q_i$
is
\vspace{-0.1cm}
\begin{equation}
\label{eqn:utility}
U(Q_i, S) = w_i \cdot \mathcal{C}(Q_i, S)
\end{equation} 
\vspace{-0.1cm}
The sum of the utility generated over $\mathcal{Q} = \{Q_i\}$ is 
\begin{equation}
\label{eqn:utility_sum}
U(\mathcal{Q}, S) = \sum_{Q_i \in \mathcal{Q}}{U(Q_i, S)}.
\end{equation}
\vspace{-0.5cm}
\end{ndef}

Our definition of utility generalizes previous metrics like the
\textit{max-subset}~\cite{Das03} used in traditional stream load
shedding literature.  \textit{Max-subset} aims to maximize the number
of output tuples, and thus can be viewed as a special case of our
definition in which each query has unit-weight.

We also note that although it is natural to define utility as a linear
function of query matches for many CEP applications
(e.g.,~\cite{wavemark, wang11}), there may exist applications where
utility can be best defined differently (e.g. a submodular function
to model diminishing returns). Considering alternative utility
functions for CEP load shedding is an area for future work.

\condcomment{\boolean{includeall}}{
\textit{mention \cite{Tatbul03} has the 
most general form of utility, which is a graph 
with arbitrary shape? Arbitrary function
however makes global optimization hard.}
}

\begin{nexp}
\vspace{-0.1cm}
\label{exp:utility}
We continue with Example~\ref{exp:query} using the event sequence $S$
and queries $Q_1$, $Q_2$ and $Q_3$ to illustrate utility.

Suppose the utility weight $w_1$ for $Q_1$ is 1, $w_2$ is 2, and $w_3$
is 3.  Since there are 2 matches of $Q_1$, $Q_2$ and $Q_3$,
respectively, in $S$, the total utility is $2 \times 1 + 2 \times 2 +
2 \times 3 = 12$.
\vspace{-0.1cm}
\end{nexp}

\vspace{-0.2cm}
\subsection{Types of resource constraints}

In this work, we study constraints on two common types 
of computing resources: CPU and memory.

\textbf{Memory-bound load shedding}.  In this first scenario, memory
is the limiting resource. Normally, arriving event data are kept in
main memory for windowed joins until time-stamp expiry (i.e., when
they are out of active windows).  During peak loads, however, event
arrival rates may be so high that the amount of memory needed to store
all event data might exceed the available capacity. In such a case not every
arriving event can be held in memory for join processing and some
events may have to be discarded.

\begin{nexp}
\label{exp:memory}
\vspace{-0.1cm}
In our running example the event sequence $S = (A_1, B_2, C_3, D_4,
E_5, A_6, B_7, C_8, D_9, E_{10} ~ ...)$, with queries $Q_1 = SEQ(A,
C)$, $Q_2 = SEQ(C, E)$ and $Q_3 = SEQ(A,$ $B, C,$ $D)$.  Because the
sliding window of each query is 5, we know each event needs to be kept
in memory for 5 units of time. Given that one event arrives in each
time unit, a total of 5 events need to be simultaneously kept in
memory.

Suppose a memory-constrained system only has memory capacity for 3
events. In this case ``shedding'' all events of type B and D will
sacrifice the results of $Q_3$ but preserves A, C and E and meets the
memory constraint. In addition, results for $Q_1$ and $Q_2$ can be
produced using available events in memory, which amounts to a total
utility of $2\times1+2\times2 = 6$. This maximizes utility,
for shedding any other two event types yields
lower utility.
\vspace{-0.1cm}

\end{nexp}

\textbf{CPU-bound load shedding}.  In the second scenario, memory may
be abundant, but CPU becomes the bottleneck.  As a result, again only
a subset of query results can be processed.

\begin{nexp}
\vspace{-0.1cm}
\label{exp:cpu}
We revisit Example~\ref{exp:memory}, but now suppose we have a CPU
constrained system. Assume for simplicity that producing each query
match costs 1 unit of CPU.  Suppose there are 2 unit of CPU available
per 5 units of time, so only 2 matches can be produced every 5 time
units.

In this setup, producing results for $Q_2$ and $Q_3$ while shedding
others yields a utility of $2\times2+2\times3 = 10$ given the events in $S$.
This is utility maximizing because $Q_2$ and $Q_3$ have the highest
utility weights.
\vspace{-0.1cm}
\end{nexp}

\textbf{Dual-bound load shedding}.
Suppose now the system is both CPU bound and memory bound
(dual-bound).

\begin{nexp}
\vspace{-0.1cm}
We continue with Example~\ref{exp:cpu}. Suppose now
due to memory constraints 3 events can be 
kept in memory per 5 time units, and in addition 
2 query matches can be produced every 5 time units
due to CPU constraints.

As can be verified, the optimal decision is to keep 
events A, C and E while producing results for $Q_1$ and $Q_2$, which
yields a utility of $2\times1+2\times2 = 6$ given the events in $S$.
Note that results for $Q_3$ cannot be produced because it needs
four events in memory while only three can fit in memory simultaneously.

\vspace{-0.1cm}
\end{nexp}

\vspace{-0.2cm}
\subsection{Types of shedding mechanisms}

In this paper, we consider two types of shedding mechanisms, an
\textit{integral load shedding}, in which certain \textit{types} of
events or query matches are discarded altogether; and a
\textit{fractional load shedding}, in which a uniform sampling is
used, such that a portion of event types or query matches is randomly
selected and preserved.

Note that both the above mentioned load-shedding mechanisms 
are relevant in an \textit{online} setting. 
That is, settings in which a shedding decision is made for the current event
before the next arriving event is processed. This is in contrast to
offline load shedding, where decisions are made after the whole
event sequence has arrived. The reason we only focus 
on online load shedding is practicality --  
most stream applications demand real-time responsiveness; an offline 
algorithm that works after the entire event sequence has arrived 
is unlikely to be practically useful.

Performance of online algorithms
is oftentimes measured against their offline counterparts
to develop quality guarantees 
like \textit{competitive ratios}~\cite{Borodin98}. However, 
we show in the following that meaningful competitive ratios
cannot be derived for \textit{any} online CEP 
load shedding algorithms.

\begin{nprop}
\vspace{-0.2cm}
\label{0}
No online CEP load shedding algorithm, deterministic or randomized,
can have competitive ratio better than $\Omega(n)$, where $n$ is the
length of the event sequence.
\vspace{-0.2cm}
\end{nprop}

The full proof of this proposition can be found in
Appendix~\ref{sec:competitive_ratio}.  Intuitively, to see why it is
hard to bound the competitive ratio, consider the following
adversarial scenario. Suppose we have a universe of $3m$ event types,
$\Sigma = \{E_i\} \cup \{E'_i\} \cup \{E''_i\}$, $i \in [m]$.  Let there be $2m$
queries $SEQ(E_i, E''_i)$ and $SEQ(E'_i, E''_i)$, $\forall i \in [m]$, each with unit utility
weight.  The stream is known to be $(e_1, e_2, ..., e_m, X)$, where
$e_i$ is either of type $E_i$ or $E'_i$ with equal probability. In addition, $X$ is drawn from the uniform distribution
on $\{E''_i: i \in [m]\}$.  Lastly, suppose the system only has
enough memory to hold two events.  The optimal offline algorithm can
look at the type of event $X$, denoted by $E''_k$, and keep the
corresponding event $e_k$ (of type $E_k$ or $E'_k$) that arrived previously, to
produce a match (either $(E_k, E''_k)$ or $(E'_k, E''_k)$, as the case may be) of utility of 1. In comparison, an
online algorithm needs to select one event  into memory before the
event type of $X$ is revealed. (Note that the offline algorithm cannot
just output results based on the last event $X$ given the form of the input,
because $e_k$ could be either $E_k$ or $E'_k$.) Thus, the probability of producing a
match is $\frac{1}{m}$, and the expected utility is also
$\frac{1}{m}$.

This result essentially suggests that we cannot hope to devise online
algorithms with good competitive ratios. In light of this result, 
in what follows, we will characterize the arriving event stream, 
and focus on optimizing the \textit{expected utility} of online
algorithms without further discussing competitive ratio bounds.

\begin{table}[t]
\vspace{-0.3cm}
\hspace{-0.2cm}
{\small
\centering
\begin{tabular}{ | c| p{7.3cm} | }  \hline
  $\Sigma$ & The set of all possible event types \\  \hline  
  $E_j$ & Event of type $j$ \\  \hline 
  $\lambda_j$ & The number of arrived events $E_j$ in a unit time (event arrival rate) \\  \hline 
  $m_j$ & The memory cost of keeping each event of type $E_j$\\  \hline  
  $\mathcal{Q}$ & The set of query patterns \\  \hline 
  $Q_i$ & Query pattern $i$ \\  \hline 
  $|Q_i|$ & The number of event types in $Q_i$ \\  \hline 
  $w_{i}$ & The utility weight associated with $Q_i$ \\ \hline 
  $n_i$ & The number of matches of $Q_i$ in a unit time\\ \hline 
  $c_i$ & The CPU cost of producing each result for $Q_i$ \\  \hline 
  $C$ & The total CPU budget\\  \hline 
  $M$ & The total memory budget \\  \hline 
  $x_j$ & The binary selection decision of event $E_j$ \\  \hline 
  $\overline{x}_j$ & The fractional sampling decision of event $E_j$ \\  \hline 
  $y_i$ & The selection decision of query $Q_i$ \\  \hline 
  $\overline{y}_i$ & The fractional sampling decision of query $Q_i$ \\  \hline 
  $p$ & The max number of queries that one event type participates in \\ \hline
  $f$ & The fraction of memory budget that the largest query consumes \\ \hline
  $d$ & The maximum number of event types in any one query  \\ \hline
\end{tabular}
\vspace{-0.2cm}
\caption{Summary of the symbols used}
\label{tab:symbols}
\vspace{-0.2cm}
}
\end{table}

\vspace{-0.2cm}
\subsection{Modeling CEP systems}
At a high level, the decision of which event or query to shed should
depend on a number of factors, including utility weights, memory/CPU
costs, and event arrival rates. Intuitively, the more important a
query is, the more desirable it is to keep constituent events in
memory and produce results of this query. Similarly the higher the
cost is to keep an event or to produce a query match, the less
desirable it is to keep that event or produce that result.  The rate at
which events arrive is also important, as it determines CPU/memory
costs of a query as well as utility it can produce.

In order to study these trade-offs in a principled way, we consider
the following factors in a CEP system.  First, we assume that the
utility weight, $w_i$, which measures the importance of query $Q_i$
installed in a CEP system, is provided as a constant.  We also assume
that the CPU cost of producing each result of $Q_i$ is a known
constant $c_i$, and the memory cost of storing each event instance of
type $E_j$ is also a fixed constant $m_j$. Note that we do not
assume uniform memory/CPU costs across different events/queries, 
because in practice event tuples
can be of different sizes. Furthermore, the arrival
rate of each event type $E_j$, denoted by $\lambda_j$, is assumed to be 
known. This is typically obtained by sampling the arriving
stream~\cite{Cherniack03, Papaemmanouil09}.  Note that characteristics
of the underlying stream may change periodically, so the sampling
procedure may be invoked at regular intervals to obtain an up-to-date
estimate of event arrival rates.

Furthermore, we assume that the ``expected'' number of matches of
$Q_i$ over a unit time, denoted by $n_i$, can also be estimated. A
simple but inefficient way to estimate $n_i$ is to sample the arriving
event stream and count the number of matches of $Q_i$ in a fixed time
period.  Less expensive alternatives also exist.  
\condcomment{\boolean{fullpaper}}
{For example, in Appendix~\ref{sec:match_estimation}, 
we discuss an analytical way to estimate $n_i$, assuming an independent 
Poisson arrival process, which is a standard assumption in the 
performance modeling literature~\cite{Lazowska84}.}  
{
For example, assuming an independent 
Poisson arrival process, which is a standard assumption in the 
performance modeling literature~\cite{Lazowska84}, the number of 
query matches can be analytically estimated using techniques similar to~\cite{he2011}. 
}
In this work we will simply treat $n_i$
as known constants without further studying the orthogonal issue of
estimating $n_i$.

Lastly, note that since $n_i$ here is the expected number of query
matches, the utility we maximize is also optimized in an expected
sense.  In particular, it is not optimized under arbitrary arrival
event strings (e.g., an adversarial input).  
While considering load shedding in such settings 
is interesting, Proposition~\ref{0} already shows that
we cannot hope to get any meaningful bounds against certain
adversarial inputs.

\begin{table}[t]
\vspace{-0.1cm}
{\small
\centering
\begin{tabular}{ | >{\centering}m{2.2cm}| m{5.5cm} | }  \hline
  \multicolumn{2}{|c|}{Approximation Ratio} \\
\hline
IMLS  & ${p}/{(1-f)}$ [Theorem~\ref{6}] \\
\hline
IMLS$^\textrm{m}$ \hspace{20 mm}(loss minimization) &  $(\frac{1}{\tau}, \frac{1}{1-\tau})$ bi-criteria approximation,
for any $\tau \in (0,1) $  [Theorem~\ref{thm:imls_bicriteria}] \\ 
\hline
ICLS & $1 + \epsilon$, for $\epsilon > 0$ [Theorem~\ref{thm:fptas}] \\
\hline
IDLS & ${p}/{(1-f)}$ [Theorem~\ref{2}] \\ 
\hline 
IDLS$^\textrm{m}$ \hspace{20 mm} (loss minimization) &  
$(\frac{1}{\tau}, \frac{1}{1-\tau}, \frac{1}{1-\tau})$ tri-criteria approximation,
for any $\tau \in (0,1) $
 [Theorem~\ref{1}] \\ 
\hline
  \multicolumn{2}{|c|}{Relative Approximation Ratio (see Definition~\ref{def:relative_approx})} \\
\hline
FMLS \hspace{20 mm} &  $1 - O\left( |\Sigma|^{-\frac{d-2}{2}} (t^2 + 1)^{-\frac{d}{2}}\right)$ where
$t = \min \left\{ \min_{E_j} \{ \frac{\lambda_j m_j}{M} \}, \frac{1}{\sqrt{|\Sigma|}} \right\}  $  [Theorem~\ref{3}] \\ 
\hline
FMLS \hspace{20 mm}(under some assumptions) &  $O\left(1 - \frac{k!}{(k-d)! k^d} \right)$ where
$k > d $  [Theorem~\ref{4}] \\ \hline
  \multicolumn{2}{|c|}{Absolute Approximation Ratio (see Definition~\ref{def:approx})} \\ \hline
FMLS & $(1 - \beta( \frac{k!}{(k-d)! k^d}) )$-approximation, 
where  $\beta = \min \left(\min_j \left\{\frac{\lambda_j m_j}{M}\right\}, 1\right)^d$,
and $k > d$ controls approximation accuracy  [Theorem~\ref{5}]. 
\\ \hline

\end{tabular}
\vspace{-0.1cm}
\caption{Summary of approximation results}
\label{tab:results}
\vspace{-0.5cm}
}
\end{table}

The symbols used in this paper are summarized in
Table~\ref{tab:symbols}, and our main approximation
results are listed in Table~\ref{tab:results}.

\vspace{-0.2cm}
\section{Related work}
\vspace{-0.1cm}
\label{sec:related}

Load shedding has been recognized as an important problem, and a large
body of work in the stream processing literature
(e.g.,~\cite{Babcock03b, Babcock03a, Das03, Gedik07, Kang03,
  Srivastava04, Tatbul03, Wei08, Wei10}) has been devoted to this
problem.  However, existing work in the context of traditional
stream processing predominantly considers the equi-join of two streaming
relations.  This is not directly applicable to CEP joins, where each
join operator typically involves multi-relational non-equi-join (on
time-stamps).  For example, the authors in~\cite{Kang03} are among the
first to study load shedding for equi-joins operators. They proposed
strategies to allocate memory and CPU resources to two joining
relations based on arrival rates, so that the number of output tuples
produced can be maximized.  Similarly, the work~\cite{Das03} also
studies the problem of load shedding while maximizing the number of
output tuples.  It utilizes value distribution of the join columns
from the two relations to produce optimized shedding decisions for
tuples with different join attribute values.

However, the canonical two-relation equi-join studied in traditional
stream systems is only a special case of the multi-relational,
non-equi-join that dominates CEP systems.  In particular, if we view
all tuples from $R$ (resp. $S$) that have the same join-attribute
value $v_i$ as a virtual CEP event type $R_i$ (resp. $S_i$), then the
traditional stream load shedding problem is captured as a very special
case of CEP load shedding we consider, where each ``query'' has
exactly two ``event types'' ($R_i$ and $S_i$), and there are no
overlapping ``event types'' between ``queries''. Because of this
equi-join nature, shedding one event has limited ramification and is
intuitively easy to solve (in fact, it is shown to be solvable
in~\cite{Das03}).  In CEP queries, however, each event type can join
with an arbitrary number of other events, and different queries use
overlapping events.  This significantly complicates the optimization
problem and makes CEP load shedding hard.

In~\cite{Babcock03a}, sampling mechanisms are proposed to implement
load shedding for aggregate stream queries (e.g., SUM), where the key
technical challenge is to determine, in a given operator tree, where
to place sampling operators and what sampling rates to use, so that
query accuracy can be maximized.  The work~\cite{Tatbul03} studies the
similar problem of strategically placing drop operator in the operator
tree to optimize utility as defined by QoS graphs. The authors
in~\cite{Srivastava04} also consider load shedding by random sampling,
and propose techniques to allocate memory among multiple operators.

The works described above study load shedding in traditional
stream systems.  The growing popularity of the new CEP model that
focuses on multi-relational non-equi-join calls for another careful
look at the load-shedding problem in the new context of CEP.

\vspace{-0.2cm}
\section{Memory-bound load shedding}
\vspace{-0.1cm}
\label{sec:memory}

Recall that in the memory-bound load shedding, 
we are given a fixed memory budget $M$,
which may be insufficient to hold all data items 
in memory. The problem is to select a subset
of events to keep in memory, such that
the overall utility can be maximized.

\subsection{The integral variant (IMLS)}
\vspace{-0.1cm}

In the integral variant of the memory-bound
load shedding problem, a binary decision,
denoted by $x_j$, is made for
each event type $E_j$, such that event instances of type
$E_j$ are either all selected and kept
in memory ($x_j = 1$), or all discarded
($x_j = 0$). The event selection decisions
in turn determine whether query $Q_i$
can be selected (denoted by $y_i$),
because output of $Q_i$ can be produced
only if all constituent event types are selected
in memory. We formulate the resulting problem as
an optimization problem as follows.
\begin{align}
\hspace{-1cm} \text{(IMLS)} \qquad{} \max & \sum_{Q_i \in \mathcal{Q}}  { n_i w_i y_i } \label{eqn:imls_obj} \\ 
 \mbox{s.t.} ~~ & \sum_{E_j \in \Sigma} {\lambda_j m_j x_j} \leq M \label{eqn:imls_knapsack} \\
 & y_i = \prod_{E_j \in Q_i}{x_j}  \label{eqn:imls_query} \\
 & y_i, x_j \in \{0, 1\}  \label{eqn:imls_binary}
\end{align}

The objective function in Equation~\eqref{eqn:imls_obj}
says that if each query $Q_i$ is selected ($y_i=1$), then
it yields an expected utility of 
$n_i w_i$ (recall that as discussed in
Section~\ref{sec:problem}, $n_i$ models
the expected number of query matches
of $Q_i$ in a unit time, while $w_i$ 
is the utility weight of each
query match). 
Equation~\eqref{eqn:imls_knapsack}
specifies the memory constraint.
Since selecting event type $E_j$ into memory consumes 
$\lambda_j m_j$ memory, where $\lambda_j$
is the arrival rate of $E_j$ and $m_j$ is 
the memory cost of each event instance of $E_j$,
Equation~\eqref{eqn:imls_knapsack} guarantees that
total memory consumption does not exceed 
the memory budget $M$.
Equation~\eqref{eqn:imls_query}
ensures that $Q_i$ can be produced
if and only if all participating events $E_j$ 
are selected and preserved in memory
($x_j=1$, for all $E_j \in Q_i$).

\vspace{-0.1cm}
\subsubsection{A general complexity analysis}
\vspace{-0.1cm}

We first give a general complexity analysis. We 
show that this shedding problem 
is NP-hard and hard to approximate
by a reduction from the \textit{Densest k-Sub-Hypergraph (DKSH)}.

\begin{nthm}
\label{10}
\vspace{-0.2cm}
The problem of utility maximizing integral memory-bound
load shedding (IMLS) is NP-hard.
\vspace{-0.1cm}
\end{nthm}

A proof of the theorem can be found in 
Appendix~\ref{sec:imls_nphard}.
We show in the following that IMLS is also hard to approximate. 

\begin{nthm}
\vspace{-0.2cm}
\label{11}
The problem of IMLS with $n$ event types 
cannot be approximated within
a factor of $2^{(\log n)^{\delta}}$, for some $\delta > 0$, 
unless 3SAT $\in$ DTIME($2^{n^{3/4+\epsilon}}$).
\vspace{-0.2cm}
\end{nthm}

This result is obtained by observing that the
reduction from DKSH is approximation-preserving.
Utilizing an inapproximability result in~\cite{Hajiaghayi06},
we obtain the theorem above (a proof is in 
Appendix~\ref{sec:imls_inapprox}).

It is worth noting that DKSH and 
related problems are conjectured to be very
hard problems with even stronger inapproximability
than what was proved in~\cite{Hajiaghayi06}. 
For example, authors in~\cite{Feige04}
conjectured that \textit{Maximum Balanced Complete 
Bipartite Subgraph (BCBS)} is $n^{\epsilon}$ hard to approximate.
If this is the case, utilizing a reduction 
from BCBS to DKSH~\cite{Hajiaghayi06}, 
DKSH would be at least $n^{\epsilon}$ hard to approximate,
which in turn renders IMLS $n^{\epsilon}$
hard to approximate given our reduction 
from DKSH.

While it is hard to solve or approximate IMLS efficiently in general,
in the following sections we look at constraints that may
apply to real-world CEP systems, and investigate
special cases that enable us to approximate or 
even solve IMLS efficiently.

\vspace{-0.1cm}
\subsubsection{A general bi-criteria approximation}
\vspace{-0.1cm}
\condcomment{\boolean{includeall}}{
\textit{how about use randomized rounding?}
}

We reformulate the integral memory-bound
problem into an alternative optimization problem
(IMLS$^{l}$) with linear constraints as follows.
\vspace{-0.1cm}
\begin{align}
 \hspace{-1cm} {\text{(IMLS$^{l}$)}} \qquad{} \max \sum_{Q_i \in \mathcal{Q}} & { n_i w_i y_i } \label{eqn:imls^l_bi_obj}  \\
  \mbox{s.t.} \sum_{E_j \in \Sigma} & {\lambda_j m_j x_j} \leq M  \notag{} \\
 & y_i \leq x_j, \forall {E_j \in Q_i}  \label{eqn:imls^l_linear} \\
 & y_i, x_j \in \{0, 1\}  \notag{} 
\end{align}

Observing that Equation~\eqref{eqn:imls_query} 
in IMLS is essentially morphed into an equivalent
Equation~\eqref{eqn:imls^l_linear}. These two
constraints are equivalent because
$y_i, x_j$ are all binary variables, $y_i$ will be 
forced to 0 if there exists $x_j = 0$ with $E_j \in Q_i$.

Instead of maximizing utility, we consider the alternative
objective of minimizing utility loss as follows.
Set $\hat{y}_i = 1 - y_i$ be the complement of $y_i$,
which indicates whether query $Q_i$ is un-selected.
We can change the utility gain maximization IMLS$^l$
into a utility loss minimization problem IMLS$^{m}$.
Note that utility gain is maximized if and only
if utility loss is minimized.
\vspace{-0.1cm}
\begin{align}
 \hspace{-1cm} {\text{(IMLS$^{m}$)}} \qquad{} \min \sum_{Q_i \in \mathcal{Q}} & { n_i w_i \hat{y}_i } \label{eqn:imls^m_bi_obj}  \\
 \mbox{s.t.} \sum_{E_j \in \Sigma} & {\lambda_j m_j x_j} \leq M  \notag{} \\
 & \hat{y}_i \geq 1 - x_j, \forall {E_j \in Q_i}  \label{eqn:imls^m_linear} \\
 & \hat{y}_i, x_j \in \{0, 1\}  \label{eqn:imls^m_integrality}
\end{align}

In IMLS$^{m}$ Equation~\eqref{eqn:imls^m_linear}
is obtained by using $\hat{y}_i = 1 - y_i$
and Equation~\eqref{eqn:imls^l_linear}.
Using this new minimization problem with 
linear structure, we prove a
bi-criteria approximation result.
Let $OPT$ be the optimal loss with budget $M$
in a loss minimization problem,
then an $(\alpha, \beta)$-bi-criteria
approximation guarantees that its solution has
at most $\alpha \cdot OPT$ loss, while uses no
more than $\beta \cdot M$ budget.
Bicriteria approximations have been 
extensively used in the context of 
resource augmentation (e.g., 
see~\cite{Pruhs04} and references therein), where the 
algorithm is augmented with extra 
resources and the benchmark is an 
optimal solution without augmentation.

\begin{nthm}
\vspace{-0.1cm}
\label{thm:imls_bicriteria}
The problem of IMLS$^{m}$
admits a $(\frac{1}{\tau}, \frac{1}{1-\tau})$ bi-criteria-approximation,
for any $\tau \in [0,1]$.
\vspace{-0.2cm}
\end{nthm}

For concreteness,
suppose we set $\tau = \frac{1}{2}$.
Then this result states that we can efficiently find a strategy
that incurs at most 2 times the optimal
utility loss with budget $M$, 
while using no more than $2M$ memory budget.

\begin{proof}
\vspace{-0.1cm}
Given a parameter $\frac{1}{\tau}$,
we construct an event selection strategy as follows.
First we drop the integrality constraint of IMLS$^{m}$ 
to obtain its LP-relaxation. We solve the relaxed 
problem to get an optimal fractional solutions
$x^*_j$ and $\hat{y}^*_i$.

We can then divide queries $\mathcal{Q}$ into two
sets, $\mathcal{Q}^a = \{ Q_i \in \mathcal{Q} | \hat{y}^*_i \leq {\tau} \}$
and $\mathcal{Q}^r = \{ Q_i \in \mathcal{Q} | \hat{y}^*_i > {\tau} \}$. Since $\hat{y}_i$ denotes whether query $Q_i$ is
un-selected, intuitively a smaller value means the query is more likely
to be accepted. We can accordingly view $\mathcal{Q}^a$ as the set
of ``promising'' queries, 
and $\mathcal{Q}^r$ as ``unpromising'' queries.

The algorithm works as follows. It preserves
every query with $\hat{y}^*_i \leq \tau$,
by selecting constituent events of $Q_i$ into memory. 
So the set of query in $\mathcal{Q}^a$ is all
accepted, while $\mathcal{Q}^r$ 
is all rejected. 

We first show that the memory consumption
is no more than $\frac{1}{1-\tau}M$. From Equation~\eqref{eqn:imls^m_linear},
we know the fractional solutions must satisfy
\vspace{-0.4cm}

\begin{equation}
x^*_j \geq \  1 - \hat{y}^*_i, \forall E_j \in Q_i
\end{equation}

In addition, we have $\hat{y}^*_i \leq \tau, \forall Q_i \in \mathcal{Q}^a$.
So we conclude
\begin{equation}
\vspace{-0.1cm}
\label{eqn:imls^m_x}
x^*_j  \geq 1 - \tau, \forall Q_i \in \mathcal{Q}^a, E_j \in Q_i
\end{equation}

Since we know $x^*_j$ are fractional solutions to IMLS$^{m}$, we have
\vspace{-0.1cm}
\begin{equation}
\label{eqn:imls^m_budget}
\sum_{E_j \in \mathcal{Q}^a}{m_j \lambda_j x_j^*} \leq \sum_{E_j \in \mathcal{Q}^a \cup \mathcal{Q}^r}{m_j \lambda_j x_j^*} = M
\end{equation}
\vspace{-0.2cm}

Here we slightly abuse the notation and use
$E_j \in \mathcal{Q}^a$ to denote 
that there exists a query 
$Q \in \mathcal{Q}^a$ such that $E_j \in Q$.

Combining~\eqref{eqn:imls^m_x} and ~\eqref{eqn:imls^m_budget},
we have
\vspace{-0.1cm}
\begin{equation}
\notag{}
\sum_{E_j \in \mathcal{Q}^a}{m_j \lambda_j (1-\tau)} \leq M
\end{equation}
\vspace{-0.3cm}

Notice that $\sum_{E_j \in \mathcal{Q}^a}{m_j \lambda_j}$ is
the total memory consumption of our rounding algorithm,
we have 
\vspace{-0.1cm}
\begin{equation}
\notag{}
\sum_{E_j \in \mathcal{Q}^a}{m_j \lambda_j } \leq \frac{M}{1-\tau}
\end{equation}
\vspace{-0.3cm}

Thus total memory consumption cannot exceed $\frac{M}{1-\tau}$.

We then need to show that the utility 
loss is bounded by a factor of $\frac{1}{\tau}$.
Denote the optimal loss of IMLS$^{m}$ as $l^*$, and the optimal
loss with LP-relaxation as $\bar{l}^*$. We then have
$\bar{l}^* \leq l^*$ because any feasible solution to
IMLS$^{m}$ is also feasible to the LP-relaxation of IMLS$^{m}$.
In addition, we know
\vspace{-0.1cm}
\begin{equation}
\notag{}
\sum_{Q_i \in \mathcal{Q}^r}{n_i w_i \hat{y}^*_i} \leq
\sum_{Q_i \in \mathcal{Q}^a \cup \mathcal{Q}^r}{n_i w_i \hat{y}^*_i} = \bar{l}^* \leq l^*
\end{equation}
\vspace{-0.3cm}

So we can obtain
\vspace{-0.2cm}
\begin{equation}
\label{eqn:imls^m_loss}
\sum_{Q_i \in \mathcal{Q}^r}{n_i w_i \hat{y}^*_i}  \leq l^*
\end{equation}
\vspace{-0.3cm}

Based on the way queries are selected, we know for every rejected query
\begin{equation}
\vspace{-0.1cm}
\label{eqn:imls^m_y}
\hat{y}^*_i \geq \tau, \forall Q_i \in \mathcal{Q}^r
\end{equation}

Combining~\eqref{eqn:imls^m_loss} and~\eqref{eqn:imls^m_y},
we get  
\begin{equation}
\vspace{-0.1cm}
\notag{}
\sum_{Q_i \in \mathcal{Q}^r}{n_i w_i \tau} \leq l^*
\end{equation}

Observing that $\sum_{Q_i \in \mathcal{Q}^r}{n_i w_i}$ is
the utility loss of the algorithm, we conclude that
\begin{equation}
\notag{}
\vspace{-0.2cm}
\sum_{Q_i \in \mathcal{Q}^r}{n_i w_i} \leq \frac{l^*}{\tau}
\end{equation}
This bounds the utility loss from optimal $l^*$ by a factor
of $\frac{1}{\tau}$, thus completing the proof.
\end{proof}

\vspace{-0.1cm}
Note that since our proof is constructive, this gives an LP-relaxation
based algorithm to achieve 
$(\frac{1}{\tau}, \frac{1}{1-\tau})$ bi-criteria-approximation
of utility loss.

\vspace{-0.1cm}
\subsubsection{An approximable special case}
\vspace{-0.1cm}

Given that memory is typically reasonably abundant 
in today's hardware setup, in this section
we will assume that the available memory capacity
is large enough such that it can 
hold at least a few number of queries. If we set 
$f = \frac{\max_{Q_i}{\sum_{E_j \in Q_i}{m_j \lambda_j}}}{M}$
to be the ratio between the memory requirement of
the largest query and available memory $M$. 
We know if $M$ is large enough, then each query uses
no more than $f M$ memory, for some $f < 1$.

In addition, denote by $p = \max_{j} |\{Q_i | E_j \in Q_i, Q_i \in \mathcal{Q}\}|$
as the maximum number of queries that 
one event type participates in.
We note that in practice there are problems in
which each event participates in a 
limited number of queries. In such cases
$p$ will be limited to a small constant.

Assuming both $p$ and $f$ are some fixed
constants, we obtain the following 
approximation result.

\condcomment{\boolean{includeall}}{
\textit{the following result seems to apply to fractional variants too,
worth mentioning? No, it doesn't actually carry to the fractional variants}
}

\begin{nthm}
\vspace{-0.2cm}
\label{6}
Let $p$ be the maximum number of queries that
one event can participate in, and $f$ be 
the ratio between the size of the largest 
query and the memory budget defined above, IMLS
admits a $\frac{p}{1-f}$-approximation.
\vspace{-0.2cm}
\end{nthm}

The idea here is to leverage the fact that the
maximum query-participation $p$ is a constant 
to simplify the memory consumption constraint,
so that a knapsack heuristic yields utility guarantees.
\condcomment{\boolean{fullpaper}}
{
In the interest of space we present 
the full proof of this theorem in 
Appendix~\ref{sec:imls_approximable}.
}
{
In the interest of space we present 
a proof of this theorem in the full version of the paper.
}

\condcomment{\boolean{includeall}}{ 
\textit{result from "A note on set union knapsack.." ,
need to include?}

\begin{nthm}
If a CEP system is $d$ event-query-sparse for some
fixed constant $d$, the problem of IMLS is $(1-e^{(-{\frac{1}{d}})})$ 
approximable.
\end{nthm}
}

\vspace{-0.1cm}
\subsubsection{A pseudo-polynomial-time solvable special case}
\vspace{-0.1cm}

We further consider the multi-tenant case where multiple
CEP applications are consolidated into one single
server or into one cloud infrastructure where the same set of 
underlying computing resources is shared across applications.

In this multi-tenancy scenario, since
different CEP applications are interested in different aspects 
of real-world event occurrences, 
there typically is no or very limited sharing 
of events across different applications
(the hospital hygiene system HyReminder and 
hospital inventory management system CIMS 
as mentioned in the Introduction, for example, have no event types
in common. So do a network intrusion detection application
and a financial application co-located in the same cloud).  
Using a hyper-graph model, a
multi-tenant CEP system can be represented 
as a hyper-graph $H$, where each event type
is represented as a vertex and each query as a hyper-edge.
If there is no sharing of event types among CEP applications,
then each connected component of $H$ corresponds to one
CEP application. Let $k$ be the
size of the largest connected component of $H$, then $k$ is essentially
the maximum number of event types used in any one 
CEP application, which may be limited to a small constant
(the total number of event 
types across multiple CEP applications is not limited).
Assuming this is the case, we have the following
special case that is pseudo-polynomial time solvable.

\begin{nthm}
\vspace{-0.2cm}
\label{7}
In a multi-tenant CEP system where each CEP tenant
uses a disjoint set of event types, if each CEP tenant
uses no more than $k$ event types, 
the problem of IMLS can be solved in time 
$O(|\Sigma||\mathcal{Q}|M 2^{k^2})$.
\vspace{-0.2cm}
\end{nthm}

Our proof utilizes a dynamic programming approach
developed for Set-union Knapsack problem~\cite{Goldschmidt06}.
The full proof of this theorem can be found in
Appendix \ref{sec:imls_ptime}.

We note that we do not assume the total number of event types across multiple 
CEP tenants to be limited. In fact,
the running time grows linearly with the total number of
event types and queries across all CEP tenants. 

Lastly, we observe that the requirement that
events in each CEP tenant are disjoint can be relaxed.
As long as the sharing of event types between different
CEP tenants are limited, such that the size of the 
largest component of $H$ mentioned above 
is bounded by $k$, the result 
in Theorem~\ref{7} 
holds.

\vspace{-0.2cm}
\subsection{The fractional variant (FMLS)}

In this section, we consider the fractional variant
of the memory-bound load shedding problem.
In this variant, instead of taking an all-or-nothing approach
to either include or exclude
\textit{all} event instances of certain types in memory, we use
a random sampling operator~\cite{Johnson05}, 
which samples \textit{some} arriving events uniformly at random 
into the memory. Denote
by $\overline{x}_j \in [0,1]$ the sampling probability for each
event type $E_j$. The fractional variant memory-bound load 
shedding (FMLS) can be written as follows.
\vspace{-0.1cm}
\begin{align}
\hspace{-1cm} \text{(FMLS)} \qquad{} \max & \sum_{Q_i \in \mathcal{Q}}  { n_i w_i \overline{y}_i } \label{eqn:fmls_obj} \\ 
 \mbox{s.t.} ~~ & \sum_{E_j \in \Sigma} {\lambda_j m_j \overline{x}_j} \leq M \label{eqn:fmls_knapsack} \\
& \overline{y}_i = \prod_{E_j \in Q_i}{\overline{x}_j}  \label{eqn:fmls_query} \\
 & 0 \leq \overline{x}_j \leq 1 \label{eqn:fmls_bound}
\end{align}

The integrality constraints are essentially dropped
from the integral version of the problem, and are
replaced by constraints in~\eqref{eqn:fmls_bound}.
We use fractional sampling variables $\overline{x}_j$
and $\overline{y}_i$ to differentiate from binary variables
$x_j, y_i$. Note that Equation~\eqref{eqn:fmls_query} 
states that the probability that a query result is produced,
$\overline{y}_i$, is the cross-product of
sampling rates of constituent events since each event is
sampled randomly and independently of each other.

\vspace{-0.1cm}
\subsubsection{A general complexity analysis}
\vspace{-0.1cm}

In the FMLS formulation, if we fold 
Equation~\eqref{eqn:fmls_query}
into the objective function 
in~\eqref{eqn:fmls_obj}, we obtain 
\begin{equation}
\max \sum_{Q_i \in \mathcal{Q}}  { n_i w_i \prod_{E_j \in Q_i}{\overline{x}_j} }
\label{eqn:fmls_obj2}
\end{equation}
This makes FMLS a polynomial optimization problem
subject to a knapsack constraint~\eqref{eqn:fmls_knapsack}. 

Since we are maximizing the objective 
function~in Equation~\eqref{eqn:fmls_obj2}, 
it is well known that if the function is 
concave, then convex 
optimization techniques~\cite{Boyd04} can be used
to solve such problems optimally. However, 
we show that except the trivial case
where each query has exactly one event 
(i.e., \eqref{eqn:fmls_obj2} becomes linear), 
in general Equation~\eqref{eqn:fmls_obj2}
is not concave.

\begin{nlem}
\vspace{-0.15cm}
\label{8}
If the objective function in
Equation~\eqref{eqn:fmls_obj2} is non-trivial
(that is, at least one query has more than one event),
then~\eqref{eqn:fmls_obj2} is non-concave.
\vspace{-0.15cm}
\end{nlem}

We show the full proof of Lemma~\ref{8}
in Appendix~\ref{sec:nonconcave}.

Given this non-concavity result, it is unlikely that
we can hope to exploit special structures of the Hessian matrix
to solve FMLS. In particular, convex optimization techniques 
like KKT conditions or gradient descent~\cite{Boyd04}
can only provide local optimal solutions,
which may be far away from the global optimal.

On the other hand, while the general
polynomial program is known 
to be hard to solve~\cite{Bellare93, Vavasis90}, 
FMLS is a special case where all coefficients 
are positive, and the constraint is a simple 
knapsack constraint. Thus the
hardness results in~\cite{Bellare93, Vavasis90}
do not apply to FMLS. We show the hardness
of FMLS by using the Motzkin-Straus theorem~\cite{Motzkin65}
and a reduction from the Clique problem.

\begin{nthm}
\vspace{-0.15cm}
\label{9}
The problem of fractional memory-bound load shedding (FMLS) is
NP-hard. FMLS remains NP-hard even if each query has
exactly two events.
\vspace{-0.15cm}
\end{nthm}

The full proof of this theorem can be found in Appendix~\ref{sec:fmls_nphard}.

\condcomment{\boolean{includeall}}{ 
\textit{still need to work out the inapprox of FMLS.
If we have inapprox of FMLS then we 
have a dichotomy for FMLS with square-free
monomials or not}
 
\textit{maybe this inapproximability is
non-existent, as it may contradict with the PTAS
result in Theorem~\ref{thm:fmls_ptas}}

\textit{ on the other hand, utility loss
minimization may be inapproximable, because
max-gain = $ 1 - \frac{1}{C*}$,
$1-gain = \frac{1}{C*}$, which is inapproximable
as per Lemma 3.1, ~\cite{deKlerk06}. The difficulty
is total utility is not 1, but much larger, which makes
it approximable.}
}

So despite the 
continuous relaxation of the decision variables 
of IMLS, FMLS is still hard to solve. However,
in the following we show that
FMLS has special structure that allows it to be solved
approximately under fairly general assumptions.

\subsubsection{Definitions of approximation}

We will first describe two definitions of 
approximation commonly 
used in the numerical optimization literature.

The first definition is similar to the approximation ratio used
in combinatorial optimization.

\begin{ndef}
\vspace{-0.1cm}
\label{def:approx}
\cite{He09} Given a maximization problem 
P that has maximum value
$v_{max}(P) > 0$. We say a polynomial
time algorithm has an absolute approximation
ratio $\epsilon \in [0,1]$, if the value found by
the algorithm, $v(P)$,
satisfies $v_{max}(P) - v(P) \leq \epsilon~{v_{max}(P)}$.
\vspace{-0.1cm}
\end{ndef}

The second notion of \textit{relative approximation ratio}
is widely used in the optimization literature
~\cite{deKlerk06, He09, Nesterov03, Vavasis90}.

\begin{ndef}
\vspace{-0.1cm}
\label{def:relative_approx}
\cite{He09} Given a maximization problem 
P that has maximum value
$v_{max}(P)$ and minimum value $v_{min}(P)$. 
We say a polynomial time algorithm has a
relative approximation ratio $\epsilon \in [0,1]$, 
if the value found by the algorithm, $v(P)$,
satisfies $v_{max}(P) - v(P) \leq \epsilon({v_{max}(P)-v_{min}(P)})$.
\vspace{-0.1cm}
\end{ndef}

Relative approximation ratio is used to bound the 
quality of solutions relative to the possible value
range of the function. We refer to this as 
$\epsilon$-\textit{relative-approximation}
to differentiate from $\epsilon$-\textit{approximation} 
used Definition~\ref{def:approx}.

Note that in both cases, $\epsilon$ indicates the
size of the gap between an approximate
solution and the optimal value. So 
smaller $\epsilon$ values are desirable
in both definitions.

\condcomment{\boolean{includeall}}{ 
\textit{this may be too complex a result to be useful}
\textit{check if ~\cite{He09} result applies to non-multi-linear
function, if so may use that result on relative bound as a more 
general result}
}

\subsubsection{Results for relative approximation bound}
\vspace{-0.1cm}
In general, the feasible region specified in
FMLS is the intersection of a unit
hyper-cube, and the region below 
a scaled simplex. We first 
normalize~\eqref{eqn:fmls_knapsack} in FMLS
using a change of variables. Let
$\bar{x}'_j = \frac{\lambda_j m_j \overline{x}_j}{M}$
be the scaled variables. We can obtain the following 
formulation FMLS'.
\vspace{-0.2cm}
\begin{align}
\hspace{-1cm} \text{(FMLS')} \qquad{} 
\max & \sum_{Q_i \in \mathcal{Q}}  { n_i w_i \prod_{E_j \in Q_i}{\frac{M}{\lambda_j m_j} \bar{x}'_j} } \label{eqn:fmls'_obj} \\ 
\mbox{s.t.} & \sum_{E_j \in \Sigma} {\bar{x}'_j} = 1 \label{eqn:fmls'_simplex} \\
 & 0 \leq \bar{x}'_j  \leq \frac{\lambda_j m_j }{M} \label{eqn:fmls'_bound}
\end{align}

Note that the inequality 
constraint~\eqref{eqn:fmls_knapsack} in FMLS is 
now replaced by an equality 
constraint~\eqref{eqn:fmls'_simplex} in FMLS'. 
This will not change the optimal value of FMLS
as long as $\sum_{E_j}{\frac{\lambda_j m_j }{M}} \geq 1$
(otherwise, although $\sum_{E_j \in \Sigma} {\bar{x}'_j} = 1$ 
is unattainable, the memory budget becomes sufficient and 
the optimal solution is trivial). This is because all coefficients 
in~\eqref{eqn:fmls_obj} are positive, pushing
any $x_i$ to a larger value will not hurt the objective value.
Since the constraint~\eqref{eqn:fmls_knapsack} 
is active for at least one global optimal point in FMLS,
changing the inequality in the knapsack constraint 
to an equality in~\eqref{eqn:fmls'_simplex} 
will not change the optimal value.

Denote by $d = \max\{|Q_i|\}$ the
maximum number of event types in 
a query. We will assume in this section
that $d$ is a fixed constant. Note that this is a 
fairly realistic assumption, as $d$ tends 
to be very small in practice
(in HyReminder~\cite{wang11}, 
for example, the longest query has 6 event 
types, so $d = 6$).
Observe that $d$ essentially
corresponds to the degree of the polynomial
in the objective function~\eqref{eqn:fmls'_obj}.

\textbf{An approximation using co-centric balls.}
Using a randomized procedure from the 
optimization literature~\cite{He09},
we show that FMLS' can be approximated
by bounding the feasible region using
two co-centric balls to obtain
a (loose) relative-approximation ratio as follows.

\begin{nthm}
\vspace{-0.1cm}
\label{3}
The problem FMLS' admits a
relative approximation ratio $\epsilon$, where
$\epsilon = 1-$ $O\left(|\Sigma|^{-\frac{d-2}{2}}( t^2 +1)^{-\frac{d}{2}}\right)$
and $ t = \min $ $\left( \min_{E_j}{\left(\frac{\lambda_j m_j}{M}\right)}, 
\frac{1}{\sqrt{|\Sigma|}}\right)$.
\vspace{-0.1cm}
\end{nthm}

A proof of
this result can be found in Appendix~\ref{sec:imls_cocentric_balls}.
Note that this is a general result that
only provides a loose relative approximation bound,
which is a function of the degree of the polynomial
$d$, the number of event types $|\Sigma|$, 
and $\frac{\lambda_j m_j}{M}$, and cannot be adjusted
to a desirable level of accuracy.

\textbf{An approximation using simplex.}
We observe that the feasible region defined in FMLS' 
has special structure. It is a subset of a simplex, which
is the intersection between a standard simplex
(Equation~\eqref{eqn:fmls'_simplex}) and
box constraints (Equation~\eqref{eqn:fmls'_bound}).

There exists techniques that produces
relative approximations for polynomials
defined over simplex. For example,~\cite{deKlerk06}
uses a grid-based approximation 
technique, where the idea is to 
fit a uniform grid into the simplex, so that values on 
all nodes of the grid are computed and the best value
is picked as an approximate solution. 
Let $\Delta_n$ be an $n$ dimensional 
simplex, then $(x \in \Delta_n  | kx \in Z_+^n) $
is defined as a $k$-grid over the simplex.
The quality of approximation is determined
by the granularity the uniform grid: the finer
the grid, the tighter the approximation bound is.

The result in~\cite{deKlerk06}, however,
does not apply here because the
feasible region of FMLS'  represents
a subset of the standard simplex. We note that there
exist a special case where if $\frac{\lambda_j m_j}{M} \geq 1, \forall j$
(that is, if the memory requirement of a single event
type exceeds the memory capacity), the feasible
region degenerates to a simplex, such that we can
use grid-based technique for approximation.

\begin{nthm}
\vspace{-0.1cm}
\label{4}
In FMLS', if for all $j$ we have $\frac{\lambda_j m_j}{M} \geq 1$ then the problem admits 
a relative approximation ratio of $\epsilon$, where
\vspace{-0.1cm}
\[
\epsilon = O \left (1-  \frac{k!}{(k-d)!k^d} \right)
\]
for any $k \in \mathbb{Z^+}$ such that $k > d$.
Here $k$ represents the number of grid points
along one dimension. 
\vspace{-0.1cm}
\end{nthm}
Note that as $k \rightarrow \infty$,
approximation ratio $\epsilon \rightarrow 0$.

The full proof of this result can be found in
Appendix~\ref{sec:fmls_relative_approx}.
This result can provide increasingly accurate
relative approximation bound for a larger $k$.
It can be shown that for a given $k$, a total of
$\binom{|\Sigma|+k-1}{|\Sigma|-1}$ number
of evaluations of the objective function is needed.

\subsubsection{Results for absolute approximation bound}
\vspace{-0.1cm}

Results obtained so far use the notion of relative approximation
(Definition~\ref{def:relative_approx}).
In this section we discuss a special case in which
FMLS' can be approximated relative to the optimal
value (Definition~\ref{def:approx}).

We consider a special case in which queries are 
\textit{regular} with no repeated events. 
That is, in each query, no events of the 
same type occur more than once
(e.g., query SEQ(A,B,C) is a query without repeated
events while SEQ(A,B,A) is not because event A appears twice). 
This may be a fairly general assumption, as queries typically
detect events of different types. HyReminder~\cite{wang11} queries,
for instance, use no repeated events in the same query
(two such example Q1 and Q2 are given in the Introduction).
In addition, we require that each query has the
same length as measured by the number of events.

With the assumption above, the objective
function Equation~\eqref{eqn:fmls'_obj} becomes a homogeneous
multi-linear polynomial, while the feasible region 
is defined over a sub-simplex that is the intersection of 
a cube and a standard simplex. We extend
a random-walk based argument in~\cite{Nesterov03}
from standard simplex to the sub-simplex,
and show an (absolute) approximation bound.

\begin{nthm}
\vspace{-0.1cm}
\label{5}
In FMLS', if $\frac{\lambda_j m_j}{M}$s are fixed constants, 
in addition if every query has no repeated event
types and is of same query length, $d$, 
then a constant-factor 
approximation can be obtained for FMLS' in 
polynomial time. 
In particular, we can achieve a 
$(1 - \beta( \frac{k!}{(k-d)! k^d}) )$-approximation, by evaluating 
Equation~\eqref{eqn:fmls_obj2} at
most $\binom{|\Sigma|+k-1}{|\Sigma|-1}$ times,
where  $\beta = \min \left(\min_j \left(\frac{\lambda_j m_j}{M}\right), 1\right)^d$,
and $k > d$ is a parameter that controls approximation accuracy.
\vspace{-0.1cm}
\end{nthm}

We use a scaling method to extend the random-walk argument in~\cite{Nesterov03}
to the sub-simplex in order to get the desirable constant
factor approximation. 
\condcomment{\boolean{fullpaper}}
{
A proof of Theorem~\ref{5} can be found in 
Appendix~\ref{sec:fmls_approx}. 
}
{
The detailed proof can be found in
the full version of the paper in the interest of space.
}
Note that, by selecting $k = O(d^2)$ we  can get
$\frac{k!}{(k-d)! k^d}$ close to $1$. Also note that
if $\frac{\lambda_j m_j}{M} \geq 1$ for all $j$,
$\beta = 1$ and we can get an 
approximation arbitrarily close to the
optimal value by using large $k$.

\vspace{-0.2cm}
\section{CPU-bound load shedding}
\vspace{-0.1cm}
\label{sec:cpu}

In this section we consider the scenario where
memory is abundant,
while CPU becomes the limiting resource that needs to be
budgeted appropriately. CPU-bound problems 
turn out to be easy to solve.

\subsection{The integral variant (ICLS)}
\vspace{-0.1cm}

In the integral variant CPU load shedding, 
we again use binary variables $y_i$ to 
denote whether results of $Q_i$ can be 
generated. For each query $Q_i$, 
at most $n_i$ number of query matches 
can be produced. Assuming the utility 
weight of each result is $w_i$, and the CPU cost 
of producing each result is $c_i$, when $Q_i$ 
is selected ($y_i = 1$) a total of $n_i w_i$ 
utility can be produced, at the same time 
a total of $n_i c_i$ CPU resources are
consumed. That yields the following ICLS 
problem.
\vspace{-0.1cm}
\begin{align}
\hspace{-1cm} \text{(ICLS)} \qquad{} \max \sum_{Q_i \in \mathcal{Q}} & { n_i w_i y_i } \label{eqn:icls_obj} \\
  \mbox{s.t.} \sum_{Q_i \in \mathcal{Q}} & {n_i c_i y_i} \leq C \label{eqn:icls_knapsack} \\
 & y_i \in \{0, 1\}  \label{eqn:icls_binary}
\end{align}

ICLS is exactly the standard 0-1 knapsack problem,
which has been studied extensively. We simply cite 
a result from~\cite[Chapter~5]{Vazirani03} for completeness.

\begin{nthm}
\vspace{-0.1cm}
\label{thm:fptas} 
\cite{Vazirani03} The integral CPU-bound load 
shedding (ICLS) is NP-complete. It admits a 
fully polynomial approximation scheme (FPTAS).
\vspace{-0.1cm}
\end{nthm}

ICLS is thus an easy variant among the load shedding
problems studied in this work.

\subsection{The fractional variant (FCLS)}
\vspace{-0.1cm}

Similar to the memory-bound load shedding problems,
we also investigate the fractional variant where 
a sampling operator is used to select a fixed fraction of
query results. 
Instead of using binary variables $y_i$,
we denote by $\overline{y}_i$ the percentage of
queries that are sampled and processed by CPU
for output. The integrality constraints in ICLS
are again dropped, and we 
can have the following FCLS formulation.
\vspace{-0.1cm}
\begin{align}
\hspace{-1cm} \text{(FCLS)} \qquad{} 
\max \sum_{Q_i \in \mathcal{Q}} & { n_i w_i \overline{y}_i } \label{eqn:fcls_obj} \\
  \mbox{s.t.} \sum_{Q_i \in \mathcal{Q}} & {n_i c_i \overline{y}_i} \leq C \label{eqn:fcls_knapsack} \\
 & 0 \leq \overline{y}_i \leq 1, \text{for all}~i \label{eqn:fcls_binary}
\end{align}

Since $n_i$, $w_i$, $c_i$ are all constants, 
this is a simple linear program problem that
can be solved in polynomial time. We 
conclude that FCLS can be efficiently solved.

\vspace{-0.3cm}
\section{Dual-bound load shedding}
\label{sec:dual}

Lastly, we study the dual-bound load shedding problem,
where both CPU and memory resources can be limited.

\vspace{-0.1cm}
\subsection{The integral variant (IDLS)}

The integral dual-bound load shedding (IDLS) can be
formulated by combining CPU and memory constraints.
\vspace{-0.1cm}
\begin{align}
\hspace{-1cm} \text{(IDLS)} \qquad{} \max & \sum_{Q_i \in \mathcal{Q}}  { n_i w_i y_i } \notag \\ 
 \mbox{s.t.} ~~ & \sum_{E_j \in \Sigma} {\lambda_j m_j x_j} \leq M \notag \\ 
& \sum_{Q_i \in \mathcal{Q}}  { n_i c_i y_i } \leq C \notag \\
& y_i \leq \prod_{E_j \in Q_i}{x_j}  \label{eqn:idls'_query} \\
 & x_j, y_i \in \{0, 1\}  \notag
\end{align}

Binary variables $x_j$, $y_i$ again
denote event selection and query selection, respectively.
Note that in order to respect the
CPU constraint, not all queries whose constituent 
events are available in memory can be produced. 
This is modeled by an inequality in
Equation~\eqref{eqn:idls'_query}.

We show that the loss minimization 
version of IDLS admits a tri-criteria approximation.

\begin{nthm}
\vspace{-0.1cm}
\label{1}
Denote by $M$ the given memory budget $C$ the given
CPU budget, and $l^*$
the optimal utility loss with that budget.
IDLS admits $(\frac{1}{\tau}, \frac{1}{1-\tau}, 
\frac{1}{1-\tau})$ tri-criteria-approximation
for any $\tau \in [0,1]$.
That is, for any $\tau \in [0,1]$, 
we can compute a strategy that uses no 
more than $\frac{1}{1-\tau}M$ memory 
$\frac{1}{1-\tau}C$ CPU, and incur no more than 
$\frac{l^*}{\tau}$ utility loss.
\vspace{-0.1cm}
\end{nthm}

The idea of the proof is to use LP-relaxation and
round the resulting fractional solution, which is 
similar to Theorem~\ref{thm:imls_bicriteria}. 
\condcomment{\boolean{fullpaper}}
{Detailed proof of Theorem~\ref{1} can be found in 
Appendix~\ref{sec:idls_tricriteria}.}
{}
In addition, we show that the approximable special 
case of IMLS (Theorem~\ref{6}) 
also holds for IDLS. 
\condcomment{\boolean{fullpaper}}
{Details of the proof can be found in 
Appendix~\ref{sec:idls_approximable}.}
{Details of the missing proof can be found in the full version of the paper.}

\begin{nthm}
\vspace{-0.1cm}
\label{2}
Let $p$ be the maximum number of queries that
one event can participate in, and 
$f = \frac{\max_{Q_i}{\sum_{E_j \in Q_i}{m_j \lambda_j}}}{M}$ be 
the ratio between the size of the largest 
query and the memory budget. IDLS
is $\frac{p}{1-f}$-approximable in 
pseudo polynomial time.
\vspace{-0.1cm}
\end{nthm}

\vspace{-0.1cm}
\subsection{The fractional variant (FDLS)}

The fractional dual-bound problem once again 
relaxes the integrality constraints in IDLS
to obtain the following FDLS problem.
\vspace{-0.1cm}
\begin{align}
\hspace{-1cm} \text{(FDLS)} \qquad{} \max & \sum_{Q_i \in \mathcal{Q}}  { n_i w_i \overline{y}_i} \label{eqn:fdls_obj} \\
 \mbox{s.t.} ~~ & \sum_{E_j \in \Sigma} {\lambda_j m_j \overline{x}_j} \leq M \notag \\
& \sum_{Q_i \in \mathcal{Q}}  { n_i c_i \overline{y}_i} \leq C \label{eqn:fdls_cpu} \\
& \overline{y}_i \leq \prod_{E_j \in Q_i}{\overline{x}_j}  \label{eqn:fdls_query} \\
 & \overline{x}_j, \overline{y}_i \in [0, 1]  \notag
\end{align}

First of all, we note that the hardness result in 
Theorem~\ref{9} 
still holds for FDLS, because 
FMLS is simply a special case of FDLS without 
constraint~\eqref{eqn:fdls_cpu}.

The approximation results established
for FMLS, however, do not carry over easily.  
If we fold~\eqref{eqn:fdls_query} into 
Equation~\eqref{eqn:fdls_obj}, and 
look at the equivalent minimization
problem by taking the inverse of the objective function,
then by an argument similar to Lemma~\ref{8}, 
we can show that objective function is
non-convex in general. In addition, 
we can show that except
in the trivial linear case, the constraint in 
Equation~\eqref{eqn:fdls_cpu}
is also non-convex using a similar argument. 
So we are dealing with a non-convex optimization
subject to non-convex constraints.

It is known that solving or even approximating
non-convex problems with
non-convex constraints to global optimality is 
hard~\cite{Boyd04, d'Aspremont03}. Techniques
dealing with such problems are relatively scarce
in the optimization literature. 
Exploiting special structure of FDLS to optimize utility
is an interesting area for future work.

\vspace{-0.2cm}
\section{Conclusions and future work}
\label{sec:conclusions}

In this work we study the problem of load shedding
in the context of the emerging Complex Event Processing
model. We investigate six variants of CEP load shedding
under various resource constraints, and demonstrate
an array of complexity results. Our results shed some
light on the hardness of load shedding CEP queries,
and provide some guidance
for developing CEP shedding algorithms in practice.

CEP load shedding is a rich problem that so
far has received little attention from the research community. 
We hope that our work will serve 
as a springboard for future research in this important
aspect of the increasingly popular CEP model.


\bibliographystyle{plain}
\bibliography{CEP_ls} 

\begin{thebibliography}{10}

\bibitem{streaminsight}
{Microsoft StreamInsight}:
  http://www.microsoft.com/sqlserver/2008/en/us/r2-complex-event.aspx.

\bibitem{streambase}
{StreamBase}: http:://www.streambase.com.

\bibitem{sybase}
{Sybase Aleri} streaming platform high-performance {CEP}:
  \\{http://www.sybase.com/files/Data\_Sheets/~Sybase\_Aleri\_StreamingPlatform\_ds.pdf}.

\bibitem{wavemark}
{Wavemark} system: Clinical inventory management solution
  www.wavemark.net/healthcare.action.

\bibitem{Adi12}
Asaf Adi, David Botzer, Gil Nechushtai, and Guy Sharo.
\newblock Complex event processing for financial services.
\newblock {\em Money Science}.

\bibitem{Agrawal08}
Jagrati Agrawal, Yanlei Diao, Daniel Gyllstrom, and Neil Immerman.
\newblock Efficient pattern matching over event streams.
\newblock In {\em proceedings of SIGMOD}, 2008.

\bibitem{Ali09}
M.~H. Ali and C.~Gerea et~al.
\newblock Microsoft cep server and online behavioral targeting.
\newblock In {\em proceedings of VLDB}, 2009.

\bibitem{Aniello11}
Leonardo Aniello, Giorgia Lodi, and Roberto Baldoni.
\newblock Inter-domain stealthy port scan detection through complex event
  processing.
\newblock In {\em Proceedings of EWDC}, 2011.

\bibitem{Babcock03a}
B.~Babcock, M.~Datar, and R.~Motwani.
\newblock Load shedding techniques for data stream systems.
\newblock In {\em Workshop on Management and Processing of Data Streams}, 2003.

\bibitem{Babcock03b}
Brian Babcock, Shivnath Babu, Rajeev Motwani, and Mayur Datar.
\newblock Chain: Operator scheduling for memory minimization in data stream
  systems.
\newblock In {\em proceedings of SIGMOD}, 2003.

\bibitem{Bellare93}
Mihir Bellare and Phillip Rogaway.
\newblock The complexity of approximating a nonlinear program.
\newblock In {\em Mathematical Programming 69}, 1993.

\bibitem{Borodin98}
A.~Borodin and R~El-Yaniv.
\newblock {\em Online Computation and Competitive Analysis}.
\newblock Cambridge University Press, 1998.

\bibitem{jm2002}
J.~M. Boyce and D.~Pittet.
\newblock Guideline for hand hygiene in healthcare settings.
\newblock {\em CDC Recommendations and Reports}, 51, 2002.

\bibitem{Boyd04}
Stephen Boyd and Lieven Vandenberghe.
\newblock {\em Convex Optimization}.
\newblock Cambridge University Press, 2004.

\bibitem{carter09}
Kim Carter.
\newblock Optimizing critical medical supplies in an acute care setting.
\newblock {\em RFID in Health Care}, 2009.

\bibitem{Chandramouli10}
Badrish Chandramouli, Jonathan Goldstein, and David Maier.
\newblock High-performance dynamic pattern matching over disordered streams.
\newblock In {\em proceedings of VLDB}, 2010.

\bibitem{Cherniack03}
Mitch Cherniack, Hari Balakrishnan, Magdalena Balazinska, Don Carney, Ugur
  Cetintemel, Ying Xin, and Stan Zdonik.
\newblock Scalable distributed stream processing.
\newblock In {\em proceedings of CIDR}, 2003.

\bibitem{chi11a}
Yun Chi, Hyun~Jin Moon, and Hakan Hacigümüs.
\newblock icbs: Incremental costbased scheduling under piecewise linear slas.
\newblock In {\em proceedings of VLDB}, 2011.

\bibitem{chi11b}
Yun Chi, Hyun~Jin Moon, Hakan Hacigümüs, and Jun'ichi Tatemura.
\newblock Sla-tree: a framework for efficiently supporting sla-based decisions
  in cloud computing.
\newblock In {\em proceedings of EDBT}, 2011.

\bibitem{Das03}
Abhinandan Das, Johannes Gehrke, and Mirek Riedewald.
\newblock Approximate join processing over data streams.
\newblock In {\em proceedings of SIGMOD}, 2003.

\bibitem{d'Aspremont03}
Alexandre d'Aspremont and Stephen Boyd.
\newblock Relaxations and randomized methods for nonconvex qcqps.

\bibitem{deKlerk06}
E.~de~Klerk, M.~Laurent, and P.~Parrilo.
\newblock A {PTAS} for the minimization of polynomials of fixed degree over the
  simplex.
\newblock In {\em Theoretical Computer Science}, 2006.

\bibitem{Diao07}
Yanlei Diao, Neil Immerman, and Daniel Gyllstrom.
\newblock Sase+: An agile language for kleene closure over event streams.
\newblock Technical report, University of Massachusetts at Amherst, 2007.

\bibitem{Feige04}
Uriel Feige and Shimon Kogan.
\newblock Hardness of approximation of the balanced complete bipartite subgraph
  problem.
\newblock Technical report, Weizmann Institute, 2004.

\bibitem{Gedik08}
Bugra Gedik, Kun-Lung Wu, and Philip~S. Yu.
\newblock Efficient construction of compact shedding filters for data stream
  processing.
\newblock In {\em proceedings of ICDE}, 2008.

\bibitem{Gedik07}
Bugra Gedik, Kun-Lung Wu, Philip~S. Yu, and Ling Liu.
\newblock A load shedding framework and optimizations for m-way windowed stream
  joins.
\newblock In {\em proceedings of ICDE}, 2007.

\bibitem{Goldschmidt06}
Olivier Goldschmidt, David Nehme, and Gang Yu.
\newblock Note: On the set-union knapsack problem.
\newblock In {\em Applied Probability \& Statistics}, 2006.

\bibitem{Gyllstrom07}
Daniel Gyllstrom, Eugene Wu, Hee-Jin Chae, Yanlei Diao, Patrick Stahlberg, and
  Gordon Anderson.
\newblock {SASE}: Complex event processing over streams.
\newblock In {\em proceedings of CIDR}, 2007.

\bibitem{Hajiaghayi06}
M.~T. Hajiaghayi, K.~Jain, K.~Konwar, L.~C. Lau, I.~I. Mandoiu, A.~Russell,
  A.~Shvartsman, and V.~V. Vazirani.
\newblock The minimum k-colored subgraph problem in haplotyping and {DNA}
  primer selection.
\newblock In {\em International Workshop on Bioinformatics Research and
  Applications}, 2006.

\bibitem{He09}
Simai He, Zhening Li, and Shuzhong Zhang.
\newblock General constrained polynomial optimization: an approximation
  approach.
\newblock In {\em Technical report}, 2009.

\bibitem{Johnson05}
Theodore Johnson, S.~Muthukrishnan, and Irina Rozenbaum.
\newblock Sampling algorithms in a stream operator.
\newblock In {\em proceedings of SIGMOD}, 2005.

\bibitem{Kang03}
Jaewoo Kang, Jeffrey~F. Naughton, and Stratis~D. Viglas.
\newblock Evaluating window joins over unbounded streams.
\newblock In {\em proceedings of ICDE}, 2003.

\bibitem{Kleiminger11}
Wilhelm Kleiminger, Evangelia Kalyvianaki, and Peter Pietzuch.
\newblock Balancing load in stream processing with the cloud.
\newblock In {\em Workshop on Self-Managing Database Systems}, 2011.

\bibitem{Lazowska84}
E.~D. Lazowska, J.~Zahorjan, G.~S. Graham, and K.~C. Sevcik.
\newblock {\em Quantitative System Performance}.
\newblock Prentice Hall, 1984.

\bibitem{Liu09}
M.~Liu, M.~Li, D.~Golovnya, E.~A. Rundensteiner, and K.~Claypool.
\newblock Sequence pattern query processing over out-of-order event streams.
\newblock In {\em proceedings of ICDE}, 2009.

\bibitem{Mei09}
Yuan Mei and Samuel Madden.
\newblock Zstream: A cost-based query processor for adaptively detecting
  composite events.
\newblock In {\em proceedings of SIGMOD}, 2009.

\bibitem{Motzkin65}
T.~Motzkin and E.~Straus.
\newblock Maxima for graphs and a new proof of a theorem of turán.
\newblock {\em Canadian Jornal of Mathamatics}, 17, 1965.

\bibitem{Narayanan11}
Sivaramakrishnan Narayanan and Florian Waas.
\newblock Dynamic prioritization of database queries.
\newblock In {\em proceedings of ICDE}, 2011.

\bibitem{Nesterov03}
Yu. Nesterov.
\newblock Random walk in a simplex and quadratic optimization over simplex
  polytopes.
\newblock {\em Center for Operations Research and Econometrics (CORE)}, 2003.

\bibitem{Papaemmanouil09}
Olga Papaemmanouil, Ugur Cetintemel, and John Jannotti.
\newblock Supporting generic cost models for wide-area stream processing.
\newblock In {\em proceedings of ICDE}, 2009.

\bibitem{Pruhs04}
K.~Pruhs~J. Sgall and E.~Torng.
\newblock {\em Handbook of scheduling: Algorithms, models, and performance
  analysis}.
\newblock 2004.

\bibitem{Srivastava04}
Utkarsh Srivastava and Jennifer Widom.
\newblock Memory-limited execution of windowed stream joins.
\newblock In {\em proceedings of VLDB}, 2004.

\bibitem{Tatbul03}
Nesime Tatbul, Ugur Cetintemel, Stanley~B. Zdonik, Mitch Cherniack, and Michael
  Stonebraker.
\newblock Load shedding in a data stream manager.
\newblock In {\em proceedings of VLDB}, 2003.

\bibitem{Vavasis90}
Stephen~A. Vavasis.
\newblock Approximation algorithms for concave quadratic programming.
\newblock Technical report, Cornell University, 1990.

\bibitem{Vazirani03}
Vijay Vazirani.
\newblock {\em Approximation Algorithms}.
\newblock Springer-Verlag Berlin Heidelberg, 2003.

\bibitem{wang11}
Di~Wang and Elke Rundensteiner.
\newblock Active complex event processing over event streams.
\newblock In {\em proceedings of VLDB}, 2011.

\bibitem{Wei08}
Mingzhu Wei, Elke~A. Rundensteiner, and Murali Mani.
\newblock Utility-driven load shedding for {XML} stream processing.
\newblock In {\em WWW}, 2008.

\bibitem{Wei10}
Mingzhu Wei, Elke~A. Rundensteiner, and Murali Mani.
\newblock Achieving high output quality under limited resources through
  structure-based spilling in {XML} streams.
\newblock In {\em proceedings of VLDB}, 2010.

\bibitem{White07}
Walker White, Mirek Riedewald, Johannes Gehrke, and Alan Demers.
\newblock What is ``next'' in event processing?
\newblock In {\em proceedings of PODS}, 2007.

\bibitem{Widder09}
Alexander Widder, Rainer v.~Ammon, Gerit Hagemann, and Dirk Schönfeld.
\newblock An approach for automatic fraud detection in the insurance domain.
\newblock In {\em AAAI Spring Symposium: Intelligent Event Processing}, 2009.

\bibitem{Wu06}
Eugene Wu, Yanlei Diao, and Shariq Rizvi.
\newblock High-performance complex event processing over streams.
\newblock In {\em proceedings of SIGMOD}, 2006.

\bibitem{Xing06}
Ying Xing, Jeong-Hyon Hwang, Ugur Cetintemel, and Stanley Zdonik.
\newblock In {\em Providing Resiliency to Load Variations in Distributed Stream
  Processing}, 2006.

\bibitem{Yao77}
Andrew Yao.
\newblock Probabilistic computations: Toward a unified measure of complexity.
\newblock In {\em proceedings of FOCS}, 1977.

\end{thebibliography}

\appendix

\condcomment{\boolean{fullpaper}}
{
\section{Additional CEP join semantics}
\label{sec:semantics}

In order to define additional CEP join
semantics, we define two additional types of subsequences.

\begin{ndef}
Given an event sequence $S = (e_{{1}},
e_{{2}}, ... e_{{N}})$ and a subsequence
$S' = (e_{i_{1}}, e_{i_{2}}, ... e_{i_{m}})$.
The subsequence $S'$ is called a
\textbf{contiguous subsequence} of $S$,
if $i_{j}+1 = i_{j+1}$, for all $ 1 \leq j \leq m-1$.

The subsequence $S'$ is called a
\textbf{type-contiguous subsequence} of $S$,
if for all events $e_{i_{j}}$ of type $E_{l_{j}}$,
where $j \in [1, m]$, there does not exist another event 
$e_k$ in $S$, where ${i_{j-1}} < {k} < {i_{j}}$,
such that $e_k$ is also of the type $E_{l_{j}}$.
\end{ndef}

We illustrate these definitions using the following example.

\begin{nexp}
\label{exp:subsequence}
Suppose there are four event types 
$\Sigma $ $= \{A, B, $ $C, $ $D\}$. 
Let the stream of arriving events be $S = $ $(A_{1}, B_{2}, $ $C_{3},
D_{4}, $ $A_{5}, B_{6}, $ $B_{7}, $ $C_{8}, D_9)$, where the 
subscript denotes the time-stamp of the event occurrence.

In this example, $(A_{1}, B_{2}, $ $C_{3})$ is a contiguous subsequence,
because intuitively there is no intervening events between $A_1$ and $B_2$, or $B_2$ and $C_3$. However,
$(A_{5}, $ $B_{6}, $ $C_{8})$ is not a contiguous subsequence,
because there is a $B_{7}$ between $B_{6}$ and $C_{8}$, violating
the definition of contiguity.

Although $(A_{5}, $ $B_{6}, $ $C_{8})$ is not a contiguous 
subsequence, it is a type-contiguous
subsequence, because there is no event of type $B$ between 
$A_{5}$ and $B_{6}$, and no event of type $C$ between 
$B_{6}$ and $C_{8}$. The subsequence 
$(A_{5}, $ $B_{7}, $ $C_{8})$, on the other hand, is not
type-contiguous subsequence because
there is an event of type $B$, $B_6$, between 
$A_{5}$ and $B_{7}$.

\end{nexp}

We are now ready to define 
\textsf{skip-till-next-match}, and \textsf{contiguity}
\footnote{An additional semantics, 
\textit{partitioned-contiguity}, was discussed 
in~\cite{Agrawal08}. Since \textit{partitioned-contiguity} 
can be simulated using \textit{contiguity} 
by sub-dividing event types
using partition criteria, for simplicity 
it is not described here.}, as follows.

\begin{ndef}
Join semantics 
\textsf{skip-till-any-match}, \textsf{skip-till-next-match},
and \textsf{contiguity} determine what event subsequence 
constitutes a \textbf{query match}.  Specifically,
a subsequence $S' = (e_{i_{1}}, e_{i_{2}},$
$...$ $, e_{i_{m}})$ of $S$ is considered 
a \textbf{query match} of $Q = SEQ(q_1, q_2, ...,$ $q_n)$ if: 

Event $e_{i_{l}}$ in $S'$ is 
of type $q^{}_{l}$ for all $l \in [1, n]$
(pattern matches), and $t_{i_{n}} - t_{i_{1}} 
\leq T(Q)$ (within window).  And in addition,
\\
(1) For \textsf{contiguity}:

The subsequence S' has to be a contiguous
subsequence.
\\
(2) For \textsf{skip-till-next-match}: 

The subsequence S' has to be a type-contiguous
subsequence.
\\
(3) For \textsf{skip-till-any-match}: 

The subsequence S' can be any subsequence.

\end{ndef}

We illustrate query matches using Example~\ref{exp:query_all}.

\begin{nexp}
\label{exp:query_all}
We revisit the event sequence used 
in Example~\ref{exp:subsequence}.
Suppose there are three queries,
\\
$Q_1$ = skip-till-any-match$(A, B, C)$,\\
$Q_1'$ = skip-till-next-match$(A, B, C)$,\\
$Q_1''$ = contiguity$(A, B, C)$, \\
where $T(Q_1)$ = $T(Q_1')$ = $T(Q_1'')$ = 5.

For $Q_1$, $(A_1, B_2, C_3)$, $(A_5, B_6, C_8)$ and
$(A_5, B_7, C_8)$ are all query matches, because
they are valid subsequences, the pattern matches
 $\{A, B, C\}$ specified in the query, and events are within
the time window 5. However, $(A_1,
B_7, C_8)$ is not a match even though pattern matches,
because the difference
in time stamp between $C_8$ and $A_1$ is over 
the window limit 5.

For $Q_1'$, $(A_1, B_2, C_3)$ and $(A_5, B_6, C_8)$ 
are query matches. Subsequence $(A_5, B_7, C_8)$ 
is not a match because it is not a type-contiguous
subsequence.

For $Q_1''$, only subsequence $(A_1, B_2, C_3)$ is
a valid match, because $(A_5, B_6, C_8)$ and 
 $(A_5, B_7, C_8)$ are not
contiguous subsequences.

\end{nexp}

}

\section{Proof of Proposition~\ref{0}}
\label{sec:competitive_ratio}

\begin{proof}
First, we show that no deterministic online 
shedding algorithm can achieve a 
competitive ratio that is independent of the
length of the event sequence. In order
to establish this, it is
sufficient to specify a distribution over event sequences  
for which the competitive ratio---defined to be the ratio of the expected utility obtained by the algorithm and the (offline) optimal utility---achieved by any deterministic algorithm must depend on the length of the sequence.  

We construct such a distribution 
as follows. Suppose we have 
a universe of $3m$ event types,
$\Sigma = \{E_i\} \cup \{E'_i\} \cup \{ E''_i\}$, $i \in [m]$.
Assume that there are $2m$ queries $SEQ(E_i, E''_i)$ and $SEQ(E'_i, E''_i)$, 
$\forall i \in [m]$, each with unit utility weight.
The event sequence is of the form $(e_1, e_2, ..., e_m, X)$, where for all $i \in [m]$,
$e_i$ is set to be either $E_i$ with probability $1/2$ or $E'_i$ with probability $1/2$. In addition, $X$ is drawn from
the uniform distribution over $\{E''_i: i \in [m]\}$.

Suppose the system can only hold two events in memory due to memory constraints. The optimal offline algorithm has a utility of 1.
This is because it can look at the
type of event $X$, denoted by $E''_k$, and 
keep the corresponding event $e_k$ which is of type
either $E_k$ or $E'_k$, and has arrived previously. This ensures that irrespective of the instantiation of the event stream one query (either $SEQ(E_k, E''_k)$ or $SEQ(E'_k, E''_k)$ match can be produced.

A deterministic online algorithm, on the
other hand, needs to select one event into 
memory before seeing the
event type of $X$. In particular, if $X$ is $E''_k$ then the deterministic algorithm is able to report the correct query (either $(E_k, E''_k)$ or $(E'_k, E''_k)$) only if it stored $e_k$.\footnote{Note that the algorithm can only report queries that have a match, i.e., it cannot report false positives. Hence, if $X$ is $E''_k$, then the algorithm cannot simply guess and declare a match, say $(E_k, E''_k)$, without any knowledge of $e_k$. Specifically, $e_k$ could instead be $E'_k$ and in such a case the declared match would be incorrect.}

Thus, if a deterministic algorithm stores $e_k$ then it succeeds only if $X$ is $E''_k$. The latter event happens with probability $\frac{1}{m}$; therefore, the probability
that a deterministic algorithm produces 
a match is $\frac{1}{m}$. Its
expected utility is also $\frac{1}{m}$
given that queries all have a unit weight.

Note that the competitive ration, i.e., the ratio of the utility produced
by the optimal offline algorithm, and an
online deterministic algorithm, is $\frac{1}{m}$. Here $m$ is the size of the
event stream minus one. Since an event stream
can be unbounded, the ratio can be arbitrarily
bad.

Next, we show that no randomized algorithm
can do any better. Using Yao's principle~\cite{Yao77},
we know that the expected utility 
of a randomized algorithm against the worst case
input stream, is no more than the expected
utility of the best deterministic algorithm against 
the input distribution. Since we know any 
deterministic algorithm against the input 
distribution constructed above is $\frac{1}{m}$,
it follows that the expected utility 
of a randomized algorithm against the worst case
input stream from this input distribution is at most
$\frac{1}{m}$, thus completing the proof.
\end{proof}

\condcomment{\boolean{fullpaper}}
{
\section{Query match estimations}
\label{sec:match_estimation}

In this section we describe a way in which the expected 
query matches of $Q_i$ over a unit time,
$n_i$, can be computed. It is not hard to image
that $n_i$ can be estimated, because a 
brute force approach is to simply sample 
the arriving event stream to count the 
number of query matches, $\hat{N}_i$ over time-window $T(Q_i)$, 
so that the number of query matches over unit time, $n_i$,
can be simply computed as $n_i = \frac{\hat{N}_i}{T(Q_i)}$.
In this section we will present an analytical approach to estimate
$n_i$ by using event arrival rates $\lambda_j$ under the
\textsf{skip-till-any-match} semantics.
Note that our discussion is only to show that there exist ways to
estimate $n_i$, so that we can treat $n_i$ as constants
in our load shedding problem formulation.


Assuming that events arrive in Possion process with 
arrival rate $\lambda_i$ (a typical assumption 
in the performance modeling 
literature~\cite{Lazowska84}), our estimation
is produced as follows. First, for each event type 
$E_j \in Q_i$, the expected number of $E_j$ occurrences 
in $T(Q_i)$, denoted as $l_j$, can be computed 
as $l_j = \lambda_j T(Q_i)$. Set $\sigma(Q_i)$ 
as the set of event types that are part 
of $Q_i$, $|Q_i|$ as the total number of events 
in $Q_i$, and $m(E_j)$ be the number of occurrences 
of event type $e_i$ in $Q_i$ (as an example, a
query $Q=(A, A, B)$ would have $\sigma(Q) = \{A, B\}$, 
$m(A) = 2$, $m(B) = 1$, and $|Q| = 3$). 
Let $L = \sum_{e_i \in \sigma(Q_i)}{l_j}$ be
the total number of occurrences of events 
relevant to $Q_i$ in time window $T(Q_i)$, then
the expected number 
of query matches of $Q_i$ in $T(Q_i)$ can
be estimated as
\begin{equation}
\label{eqn:count_estimation}
\tbinom{L}{|Q_i|} \frac{\prod_{e_i \in \sigma(Q_i)}{\prod_{k=0}^{m(e_i)}{(l_i - k)}}}{\prod_{r=0}^{|Q_i|}{(L-r)}}
\end{equation}

The formula of above is obtained as follows. 
Given a total of $L$ event 
occurrences in $T(Q_i)$, we only pick a total 
$|Q_i|$ events to form a query match. Let us pick 
the first $|Q_i|$ events to form random permutations 
and compute the probability that the first $|Q_i|$ 
events produces a match. Let the first position of 
$Q_i$ be an event of type $E_{p_{1}}$. The 
probability of actually seeing the event that type 
is $\frac{l_{E_{p_{1}}}}{L}$. For the second event 
in $Q_i$, if it is also of type $E_{p_{1}}$, the 
probability of seeing that event that type is 
$\frac{l_{E_{p_{1}}}-1}{L-1}$, otherwise it is 
$\frac{l_{E_{p_{1}}}}{L-1}$, so on and so forth. 
In the end we have
$\frac{\prod_{E_j \in \sigma(Q_i)}{\prod_{k=0}^{n(E_j}{(l_i - k)}}}{\prod_{r=0}^{|Q_i|}{(L-r)}}$. 
Given that there are a total of $\tbinom{L}{|Q_i|}$ 
such possible positions out of $L$ event 
occurrences and each of which is symmetric, 
the expected count of query matches can be 
expressed as the product of the two, thus 
the Equation~\eqref{eqn:count_estimation}. 
Given that over time $T(Q_i)$ this many number
of event matches can be produced, we can conclude
that over unit time the expected number of 
query matches $n_i$, is thus
\begin{equation}
n_i = \frac{1}{T(Q_i)} \tbinom{L}{|Q_i|} \frac{\prod_{e_i \in \sigma(Q_i)}{\prod_{k=0}^{m(e_i)}{(l_i - k)}}}{\prod_{r=0}^{|Q_i|}{(L-r)}} \notag
\end{equation}
}

\section{Proof of Theorem~\ref{10}}
\label{sec:imls_nphard}

\begin{proof}
We obtain the hardness result by a reduction from
\textit{Densest k-sub-hypergraph (DKSH)}. Recall that given 
a hypergraph $G=(V, H)$, the decision version of
DKSH is to determine if there exists an induced
subgraph $G'=(V', H')$, such that 
$|V'| \leq k$, and $|H'| \geq B$ for some given
constant $B$.

Given any instance of the DKSH
problem, we construct an IMLS problem as follows.
We build a bijection $\phi: V \rightarrow \Sigma$, so that each vertex 
$v_j \in V$ corresponds to one event type $E_j \in \Sigma$.
For each hyperedge $h_i \in H$, using
vertices $v_k$ that are endpoints of $h_i$ 
to construct a corresponding query $Q_i$, such that 
$\forall v_k \in h_i$, $\phi(v_k) \in Q_i$. Set all $E_j$ to have unit
cost $(m_j \lambda_j = 1)$ and all $Q_i$ have unit weight
$(w_i n_i = 1)$, and lastly set the memory budget
to $k$.

We first show the forward direction, that is if there
exists a solution to DKSH, i.e., a $k$-sub-hypergraph 
with at least $B$ hyper-edges, 
then there exists a solution to IMLS with no
more than $k$ memory cost but achieves $B$ utility. 
Suppose the subgraph $G'=(V', H')$
is the solution to DKSH. In the corresponding IMLS
problem, if we keep all event types 
$\phi(v'), \forall v' \in V'$, our solution has a memory
cost of $|V'|$, which is no more than $k$ since
$G'$ is a k-sub-hypergraph. This ensures
the constructed solution is feasible. Furthermore,
the utility of the IMLS solution is exactly $|H'|$
by our unit weight construction. Since we know $|H'| \geq B$,
this completes the proof in this direction.

In the other direction,
let the set of events selected in IMLS be 
$S \subseteq \Sigma$ while the set of dropped 
events be $D = \Sigma \setminus S$. The set of vertices $V_s$
corresponding to $S$ induces a subgraph $G_s = (V_s, H_s)$.
S respects memory-bounds in IMLS implies 
$|V_s| \leq k$, so $G_s$ is a valid $k$-sub-hyper-graph. 
In addition, S produces utility $B$ in IMLS ensures
that $|H_s| \geq B$, thus completing the hardness proof.
\end{proof}

\section{Proof of Theorem~\ref{11}}
\label{sec:imls_inapprox}

\begin{proof}
We note that the reduction 
from~\textit{Densest k-sub-hypergraph (DKSH)}
discussed above
is approximation preserving. Utilizing a
hardness result of DKSH~\cite{Hajiaghayi06},
which establishes that DKSH cannot be approximated
within a factor of $2^{(\log n)^{\delta}}$
for some $\delta > 0$, we obtain the
inapproximability result of IMLS.
\end{proof}

\condcomment{\boolean{fullpaper}}
{
\section{Proof of Theorem~\ref{6}}
\label{sec:imls_approximable}

\begin{proof}
We modify the IMLS problem in Equation~\eqref{eqn:imls_obj}-\eqref{eqn:imls_binary} 
to a relaxed version IMLS' as follows.

\begin{align}
\hspace{-1cm} \text{(IMLS')} \qquad{} 
\max & \sum_{Q_i \in \mathcal{Q}}  { n_i w_i y_i }  \notag \\ 
 \mbox{s.t.} ~~ & \sum_{Q_i}{ (y_i \sum_{E_j \in Q_i} {\lambda_j m_j }) } \leq M \label{eqn:imls'_knapsack} \\
 & y_i = \prod_{E_j \in Q_i}{x_j} \notag \\
 & x_j, y_i \in \{0, 1\} \notag
\end{align}

We essentially replace Equation~\eqref{eqn:imls_knapsack} and 
Equation~\eqref{eqn:imls_query} in IMLS with
Equation~\eqref{eqn:imls'_knapsack}. Note that
the left side of Equation~\eqref{eqn:imls'_knapsack}
is an overestimate of the memory consumption of a particular
event/query selection strategy. First, set 
$x_j^*, y_i^*$ as an optimal solution to IMLS, where
each event selected participates at least one selected query. 
That is, if we set 
$X = \{x_j^* | x_j^* = 1, \nexists Q_i \ni E_j $, $y_i^* = 1\}$, 
we have $X = \emptyset$.
There always exists one such optimal solution, because
otherwise we can force $x_j^* = 0, \forall x_j^* \in X$ without
changing the optimal value $opt^*$ (by definition
no events $E_j$ corresponding to $x_j^* \in X$ participates in
queries $Q_i$ with $y_i^* = 1$), thus obtaining one optimal
solution with $X = \emptyset$. With this $x_i^*, y_j^*$
we can show the following inequality

\begin{equation}
\sum_{E_j \in \Sigma} {\lambda_j m_j x_j^*} \leq
\sum_{Q_i}{ (y_i^* \sum_{E_j \in Q_i} {\lambda_j m_j }) } \label{eqn:ineq1}
\end{equation}

This is the case because we know $X = \emptyset$, so 
$\forall x_j^* = 1, \exists Q_i \ni E_j $ where $y_i^* = 1$.
It ensures that for each term 
${\lambda_j m_j x_j^*}$ where $x_j^* = 1$
from the left side of Equation~\eqref{eqn:ineq1}, there
exists one matching term
 $y_i^* \lambda_j m_j$  on the right side, where 
$E_j \in Q_i, y_i^* = 1$, thus providing the inequality.

This inequality guarantees that any feasible solution of IMLS'
will be feasible for IMLS. This is because given 
Equation~\eqref{eqn:imls'_knapsack} in IMLS' and 
Equation~\eqref{eqn:ineq1}, we know 
Equation~\eqref{eqn:imls_knapsack} in IMLS holds, so
the solution to IMLS' will produce the same value
in IMLS but still respect the constraints in IMLS.

In addition, we show the following inequality is true
\begin{equation}
\frac{1}{p} \sum_{Q_i}{ (y_i^* \sum_{E_j \in Q_i} {\lambda_j m_j }) } 
\leq \sum_{E_j \in \Sigma} {\lambda_j m_j x_j^*} \label{eqn:ineq2}
\end{equation}
Recall that we know each event $E_j$ participates at most
$p$ number of queries. If each event participates exactly
one query, we know 
$ \sum_{Q_i}{ (y_i^* \sum_{E_j \in Q_i} {\lambda_j m_j }) } 
= \sum_{E_j \in \Sigma} {\lambda_j m_j x_j^*} $.
Now if each event appears in no more than $p$ queries,
then for all $j$, the coefficients of each 
term $\lambda_j m_j$ on the right hand 
side is no more than $p$, thus the inequality 

With this inequality we further define the program IMLS''.

\begin{align}
\hspace{-1cm} \text{(IMLS'')} \qquad{} 
\max & \sum_{Q_i \in \mathcal{Q}}  { n_i w_i y_i }  \notag \\ 
 \mbox{s.t.} ~~ & \frac{1}{p} \sum_{Q_i}{ (y_i^* \sum_{E_j \in Q_i} {\lambda_j m_j }) } \leq M \label{eqn:imls''_knapsack} \\
 & y_i = \prod_{E_j \in Q_i}{x_j} \notag \\
 & x_j, y_i \in \{0, 1\} \notag
\end{align}

Solution feasible to IMLS will be feasible to IMLS'', because
we only replace Equation~\eqref{eqn:imls_knapsack} in IMLS
with Equation~\eqref{eqn:imls''_knapsack}, and we know 
Equation~\eqref{eqn:ineq2} holds. Let the optimal value to
IMLS'' be $opt''^*$, the optimal value of IMLS $opt^*$,
then 
\begin{equation}
opt''^* \geq opt^* \label{eqn:opt0}
\end{equation}

Furthermore, observe that both IMLS' and IMLS'' are now
simple one dimensional knapsack problems, where IMLS''
has $p$ times more knapsack capacity than IMLS'
(because we can rewrite Equation~\eqref{eqn:imls''_knapsack}
as $\sum_{Q_i}{ (y_i^* \sum_{E_j \in Q_i} {\lambda_j m_j }) } \leq p M$). Suppose
we solve IMLS' using a simple knapsack heuristics, by picking
queries (setting $y_i = 1$) in the descending 
order of $\frac{n_i w_i}{\lambda_j m_j}$ until the knapsack
$M$ is filled, and denote the value so produced by $val'$. 
We can show $val'$ is good relative to $opt''^*$ in the following sense.

\begin{equation} 
val' \times \frac{p}{1-f} \geq opt''^* \label{eqn:heuristics}
\end{equation}

To see why this is the case, suppose in the knapsack in IMLS',
we cheat by allowing each query to be divided 
into fractional parts. The optimal value
in this modified version, $\hat{opt}'$, can be produced by simply
packing queries in the order of $\frac{n_i w_i}{\lambda_j m_j}$.
The difference between $\hat{opt}'$ and the heuristics $val'$ 
is the largest when the last query cannot fit in its entirety 
using IMLS' heuristics. Since we know each query is relatively
small compared to the budget $M$ (bounded by $f M$),
we know
\begin{equation}
\frac{val'}{\hat{opt'}} \geq \frac{1-f}{1} = 1- f \label{eqn:opt1}
\end{equation}

Next, we know if we cheat again by allowing fractional 
queries in IMLS'', the optimal value $\hat{opt}''$ is no more
than $p$ times $\hat{opt}'$, 
\begin{equation}
\hat{opt}'' \leq p \times \hat{opt}' \label{eqn:opt2}
\end{equation}
because knapsack size has grown
$p$ times but we fill knapsacks strictly based on their 
$\frac{n_i w_i}{\lambda_j m_j}$ values.

Lastly, it is easy to see that 
\begin{equation}
\hat{opt}'' \geq opt''^* \label{eqn:opt3}
\end{equation}

Combining Equation~\eqref{eqn:opt1}~\eqref{eqn:opt2}~\eqref{eqn:opt3}
we can derive Equation~\eqref{eqn:heuristics}. Since we also know
Equation~\eqref{eqn:opt0}, we get
\begin{equation} 
val' \times \frac{p}{1-f}\geq opt^* \label{eqn:quality}
\end{equation}
thus completing the proof.
\end{proof}
}

\section{Proof of Theorem~\ref{7}}
\label{sec:imls_ptime}
\begin{proofsketch}
The IMLS problem studied in this work can be formulated
as a Set-union Knapsack problem~\cite{Goldschmidt06}.
A dynamic programming algorithm is proposed 
 in~\cite{Goldschmidt06} for Set-union Knapsack, which 
however runs in exponential time. If we define
an \textit{adjacency graph} $G$ by representing all
events as graph vertices, and each pair of vertices are 
connected if the corresponding events co-occur in
the same query. The exponent of the running 
time is shown to be no more than the cut-width of 
the induced adjacency graph $G$, $cw(G)$.
Recall that cut-width of a graph $G$ is defined as the smallest
integer $k$ such that the vertices of $G$ can be arranged
in a linear layout $[v_1, ..., v_n]$ such that for every $i \in [1, n-1]$,
there are at most $k$ edges with one endpoint 
in $\{v_1, ..., v_i\}$ and another endpoint in $\{v_{i+1}, ..., v_n\}$.

In the context of a multi-tenant CEP system, assuming
each non-overlapping CEP tenant uses at most
$k$ event types, then the size of the largest component
of the adjacency graph $G$ is at most 
$k$. This, combines with the fact that
the degree of each vertex in each component 
is at most $k$, ensures that $cw(G) \leq k^2$.
The running time of the dynamic
programming approach in~\cite{Goldschmidt06}
can then be bounded by
$O(|\Sigma||\mathcal{Q}|M 2^{k^2})$.
Note that this result is pseudo-polynomial, because
the running time depends on the value 
of memory budget $M$ instead of the number of bits
it needs to represent it.
\end{proofsketch}

\section{Proof of Lemma~\ref{8}}
\label{sec:nonconcave}
\begin{proof}
Denote $H$ as the Hessian
matrix of~\eqref{eqn:fmls_obj2}
representing its second order partial derivatives.
In the trivial case where~\eqref{eqn:fmls_obj2} is
just a linear function (each query has
exactly one event type), $H$ is an all-zero matrix
with all-zero eigenvalues, which is trivially concave.

In general,~\eqref{eqn:fmls_obj2}
is a nonlinear polynomial (i.e., at least
one query has more than one event).
Since~\eqref{eqn:fmls_obj2} is
a polynomial with positive coefficients
and positive exponents, and
$\overline{x}_j \geq 0, \forall j$, we know all non-zero
second order partial derivatives 
of~\eqref{eqn:fmls_obj2} are positive, and thus all
non-zero entries of $H$ are positive.
Denote by $h_{ij}$ the $i$th row, 
$j$th column entry of $H$, 
given that $h_{ij} \geq 0, \forall i, j$,
we know the trace of the Hessian matrix 
$tr(H) = h_{11} + h_{22} + ... + h_{|\Sigma||\Sigma|} \geq 0$. 
From linear algebra, we know that 
$tr(H)$ equals the sum of the eigenvalues 
of $H$.  

Since $H$ 
is a Hessian matrix, we know it is symmetric and
its eigenvalues are all real. In addition,
at least one eigenvalue is non-zero because $H$
is not an all-zero matrix. 

We show by contradiction that $H$ must have at
least one positive eigenvalue. Suppose this is not
the case. Since $H$ has at least one non-zero 
eigenvalue, all its non-zero eigenvalues have to be
negative, which implies that the sum of all eigenvalues
are negative, thus we have $tr(H) < 0$. This
contradicts with the fact that $tr(H) \geq 0$. Therefore
$H$ must have at least one positive eigenvalue,
which ensures that $H$ is non-concave. 
\end{proof}

\section{Proof of Theorem~\ref{9}}
\label{sec:fmls_nphard}

\begin{proof}
We show the hardness of this problem by a reduction
from the Clique problem. 
Given a graph $G = (V, E)$, the decision version of 
the Clique problem is to determine if there exists a
clique of size $k$ in $G$. 

From any instance of the Clique problem, we construct 
an instance of the FMLS problem as follows.
We build a bijective function $\phi: V \rightarrow \Sigma$
to map each vertex $v_j \in V$ to an event $\phi(v_j) \in \Sigma$.
We set a unit memory consumption for each event
($\lambda_j m_j = 1$), and a unit
knapsack capacity. So we get
$\sum_{E_j \in \Sigma}{ \overline{x}_j} \leq 1$.
Furthermore, given an edge $e_i = (v_{l}, v_{k}) \in E$, 
we build a query 
$(\phi(v_{l}), \phi(v_{k}))$ with unit utility weight 
($n_i w_i = 1$). 
We then essentially have a unit-weight, unit-cost, 
length-two-query FMLS that corresponds to the graph $G$.
This gives rise to the following bilinear optimization 
problem subject to a knapsack constraint.

\begin{align}
 \max & \sum_{Q_i:(E_l, E_k) \in \mathcal{Q}}  { \overline{x}_{l} \overline{x}_{k} } \notag \\ 
 \mbox{s.t.} ~~& \sum_{E_j \in \Sigma} {\overline{x}_j} \leq 1 \label{eqn:clique_knapsack} \\
 ~~ & 0 \leq \overline{x}_j \leq 1  \notag
\end{align}


Since we are dealing with a maximization problem,
and the coefficients of the objective are all non-negative, 
increasing $\overline{x}_j$ values will not ``hurt'' 
the objective. Thus, the knapsack constraint
in~\eqref{eqn:clique_knapsack}
can be changed into a standard simplex constraint
$\sum_{E_j \in \Sigma} {\overline{x}_j} = 1$ without
changing the optimal value of the problem.

Given the FMLS defined above, the decision version
of FMLS is to decide if there exists a fractional strategy 
such that the total utility value is at least $u$.
We first show that if there exists a clique of size $k$
in $G$, then the value of the FMLS
we construct is at least $\frac{k-1}{2k}$.
To show this connection, we use 
the Motzkin-Straus theorem~\cite{Motzkin65},
which states that global maximum over the standard
simplex is attained when values are distributed evenly
among the largest clique from the graph. So 
if there is a clique of size $k$ in graph
$G$, then the $k$-clique has a total of $\binom{k}{2}$ 
number of edges. Since each edge produces a value of
$(\frac{1}{k})^2$, the the optimal value of FMLS is at least 
$\binom{k}{2} (\frac{1}{k})^2$ = $\frac{k-1}{2k}$.

We now show the other direction, that is if the
optimal value FMLS problem we construct is no
less than $\frac{k-1}{2k}$, then the graph $G$ has
a clique of size no less than $k$. We show this by
contradiction. Suppose the size of the 
maximum clique of $G$ is $c$, $c < k$.
Then by the Motzkin-Straus theorem~\cite{Motzkin65}
the global maximum of the FMLS we
construct is at most $\frac{c-1}{2c}$,
which is less than $\frac{k-1}{2k}$ because 
 $c < k$. This contradicts with the fact
that the optimal value is no less than 
$\frac{k-1}{2k}$, therefore, the graph must
have a clique of size at least $k$.

We note that by using a reduction from 
Clique, we have shown that even 
in the restricted case where the 
objective function in Equation~\eqref{eqn:fmls_obj2} is 
a bi-linear function (a summation of quadratic 
square-free monomials),
or in other words each query has exactly two
events, FMLS remains NP-hard.
\end{proof}

\section{Proof of Theorem~\ref{3}}
\label{sec:imls_cocentric_balls}

\begin{proof}
Let $B(t) = \{x \in \mathcal{R}^n, \|x\|_2 \leq t \}$ be a ball
constraint. We show that the feasible region of FMLS',
denoted by S, satisfies 
$B(t) \subseteq S \subseteq B(1)$, where 
$t = \min( \min_{E_j}(\frac{\lambda_j m_j}{M}), $ $
\frac{1}{\sqrt{|\Sigma|}})$.

First, we know that any feasible solution satisfies 
$\|x\|_1 \leq 1$. It can be shown that 
$\forall x, \|x\|_2 \leq \|x\|_1$.
Thus we know $\|x\|_2 \leq 1$, 
and $S \subseteq B(1)$. 

On the other hand,
the ball inside $S$ is limited by the shortest 
edge of the hyper-rectangle,
$\min_{E_j}(\frac{\lambda_j m_j}{M})$, and
the largest possible ball inside standard simplex,
which has a radius of $\frac{1}{\sqrt{|\Sigma|}}$. 
So we have 
\begin{align*} t = \min( \min_{E_j}(\frac{\lambda_j m_j}{M}), 
\frac{1}{\sqrt{|\Sigma|}}).\end{align*}
The largest ball inside $S$ is thus
$B(t)$.

Authors in~\cite{He09} show that if a convex
feasible region
contains a ball constraint, and in addition is bounded by
another ball constraint, the polynomial program
can be approximated with a relative approximation
ratio that is a function of the degree of the polynomial,
and the radius of the ball constraints, namely the ratio is
$1-(d+1)!(2d)^{-2d}$ $(|\Sigma|+1)^{-\frac{d-2}{2}}( t^2 +1)^{-\frac{d}{2}}$. 
Given that $d$ is assumed to be a fixed constant, thus
our result in the theorem.
\end{proof}

\section{Proof of Theorem~\ref{4}}
\label{sec:fmls_relative_approx}

\begin{proof}
We again use the FMLS' formulation.
Let $d=\max\{|Q_i|\}$ be the maximum number of
events in any query. Monomials in~\eqref{eqn:fmls'_obj}
can be homogenized to degree $d$ using the fact
that $\sum_{E_j}\bar{x}'_j = 1$. So~\eqref{eqn:fmls'_obj}
is equivalent to the following formula that is degree-$d$
homogeneous
\begin{equation}
\sum_{Q_i \in \mathcal{Q}}  { n_i w_i  (\sum_{E_j}{(\bar{x}'_j)})^{d-|Q_i|} \prod_{E_j \in Q_i}{\frac{M}{\lambda_j m_j} \bar{x}'_j} }
\end{equation}

Under the assumption that for all $j$ we have $\frac{\lambda_j m_j}{M} \geq 1$, the problem then is to optimize a degree-$d$
homogeneous polynomial program on the standard simplex $\Delta_{|\Sigma|}$.

Authors in~\cite{deKlerk06} show that 
using a uniform $k$-grid, 
$\Delta_{|\Sigma|}(k) = \{ x \in \Delta_{|\Sigma|} | k x \in \mathbb{Z^+}\} $,
polynomials of fixed degree $d$
can be approximated by evaluating~\eqref{eqn:fmls'_obj}
for all $\binom{|\Sigma|+k-1}{|\Sigma|-1}$ number
of points on the k-grid $\Delta_{|\Sigma|}(k)$.
Utilizing approximation bound obtained in~\cite{deKlerk06}
(see Theorem 1.3 in~\cite{deKlerk06}),
FMLS' in this special case can be approximated
with a relative approximation ratio 
$\epsilon = (1-\frac{k!}{(k-d)!k^d}) \binom{2d-1}{d} d^d$.
Given that $d$ is assumed to be a fixed constant,
we get the result stated in the Theorem.
\end{proof}

\condcomment{\boolean{fullpaper}}
{
\section{Proof of Theorem \ref{5}}

It was shown in~\cite{Nesterov03}
that if the feasible region is the standard
simplex, a random-walk based argument can be used
to prove an (absolute) approximation bound of
a grid-based approach. Although this result does
not apply to the problem we have at hand since
our feasible region is sub-simplex instead of simplex,
we extend the random-walk argument in~\cite{Nesterov03} to
sub-simplex, and obtain a similar approximation ratio
as follows.

\label{sec:fmls_approx}

\begin{proof}
We again use the FMLS' formulation in 
Equation~\eqref{eqn:fmls'_obj}
- \eqref{eqn:fmls'_bound}. We note that
the feasible region of FMLS' is the intersection
of a standard simplex (Equation~\eqref{eqn:fmls'_simplex}),
and a set of box constraints (Equation~\eqref{eqn:fmls'_bound}).
In addition, since we know that the CEP queries 
are regular, that is, there is no repeated events in the
same query, and each query has the same number of 
events, we can conclude that the objective
function~\eqref{eqn:fmls'_obj}  is a multi-linear, 
homogeneous function.

Although there are PTAS results
for multi-linear, homogeneous function defined over
standard simplex~\cite{Nesterov03}, these results
cannot be applied to our FMLS' problem because 
our feasible region is a subset the simplex.

In~\cite{Nesterov03}, in order to obtain an approximation on the simplex, $\Delta_{|\Sigma|}$, an exhaustive search is performed over the uniform grid defined as $\Delta_{|\Sigma|}(k) := \{ x \in \Delta_{|\Sigma|} \mid kx \in \mathbb{Z}_+^{|\Sigma|} \}$. This uniform grid contains $\binom{|\Sigma|+k-1}{|\Sigma|-1}$ points, hence the algorithm performs at most $\binom{|\Sigma|+k-1}{|\Sigma|-1}$ function evaluations to determine an optimal point on the grid. 

In the following we extend the result 
in~\cite{Nesterov03}, by showing that a similar 
algorithm exists for multi-linear, homogeneous polynomial
with non-negative coefficients, defined over the intersection
of a standard simplex and a set of box constraints. 

Denote by $d$ the degree of the objective function
(or, equivalently, the number of events in each query).
Furthermore, we use $\tau_j$ to denote $\frac{\lambda_j m_j }{M}$
in~\eqref{eqn:fmls'_bound} for succinctness.
Note here, $\tau_j \leq 1, \forall j$, because otherwise given the
simplex constraint~\eqref{eqn:fmls'_simplex} and the
fact that all $x_j$ are positive, we can replace $\tau_j$ with $\tau_j > 1$
in $0 \leq x_j \leq \tau_j$ as $0 \leq x_j \leq 1$.

Similar to~\cite{Nesterov03}, first we define $p \in \Delta_n$
be a vector in the standard simplex. Denote by $\zeta(p)$ a 
discrete random variable distributed as $Pr\{\zeta(p) = i\}
= p^{(i)}$.

As in~\cite{Nesterov03}, we define a random process $x$ as
\[x_0(p) = 0 \in R^n\]
\[x_{k+1}(p) = x_k(p) + e_{\zeta_k(p)}, k \geq 0\]
In addition, we define a ``scaled-down'' version $x'$ of $x$
to reflect the box constraint~\eqref{eqn:fmls'_bound}.
\[x'^{(i)}_k(p) = x^{(i)}_{k}(p) \tau_{i}\]

Note that for all $k$, any realization $\frac{x_k(p)}{k}$ lies on the uniform grid $\Delta_{|\Sigma|}(k)$. 
Since $x'$ is a linearly scaled version of $x$, we
can use results in 
Equation~(2.5) and (2.6) of~\cite{Nesterov03},
to obtain mean, variance, and covariance of $x'$ as follows.
\[ \mu'^{(i)}_{k}(p) = E[x'^{(i)}_{k}(p)] = k p^{(i)} \tau_i \] 
\[ E\left[{(x'^{(i)}_{k}(p) - \mu_{k}'^{(i)})}^{2}\right] = k p^{(i)}(1-p^{(i)})  \tau_i^2\]
\[ E\left[ (x'^{(i)}_{k}(p) - \mu_{k}'^{(i)}) \cdot (x'^{(j)}_{k}(p) - \mu_{k}'^{(j)}) \right] = -k p^{(i)}p^{(j)} \tau_i \tau_j\]

Similarly, we can reuse results of $E[x^\gamma_k(p)]$ in Lemma 2
of~\cite{Nesterov03}. In particular, we have
\begin{equation} 
\label{eqn:nestrov_aux}
E^\gamma_k(p)' = E[x'^\gamma_k(p)]  = E[x^\gamma_k(p)] \tau^\gamma 
\end{equation}

Now we need to show that the approximation result in Lemma 3
of~\cite{Nesterov03} still holds. To be consistent with the minimization
problem studied in~\cite{Nesterov03}, we get the inverse of 
\eqref{eqn:fmls'_obj}, and consider the corresponding minimization
version of the IMLS'.

Let the objective function under consideration be $f(x) := \sum_{\alpha \in A} f_\alpha x^\alpha $, where $\alpha$ is a multi-index in $\mathbb{Z}_+^{|\Sigma|}$. Also, say $f'^*_k$ is the minimum value attained among all grid points as defined by
the random walk $x'(p)$, where $p$ is set to the optimal vector
in our feasible region~\eqref{eqn:fmls'_simplex} and~\eqref{eqn:fmls'_bound}
that minimizes $f$. We know $f'^*_k$ is no more than
the expectation of the function value over the random walk. 
\begin{equation}
\label{eqn:nesterov_step1}
f'^*_k \leq E[f(\frac{1}{k}x'_k(p))] 
\end{equation}

Since $f$ is degree $d$ homogeneous, we have
\begin{equation}
\label{eqn:nesterov_step2}
  E[f(\frac{1}{k}x'_k(p))] = \Sigma_{\alpha \in A}{f_{\alpha}E[\frac{1}{k^d}x'^\alpha_k(p)]} 
\end{equation}

Utilizing a result from Lemma 2 of~\cite{Nesterov03}
and Equation~\eqref{eqn:nestrov_aux}, we get
\begin{equation}
\label{eqn:nesterov_step3}
 \Sigma_{\alpha \in A}{f_{\alpha}E[\frac{1}{k^d}x'^\alpha_k(p)]} = \frac{k!}{(k-d)!k^{d}} \Sigma_{\alpha \in A}{f_{\alpha} p^{\alpha} \tau^{\alpha}}
\end{equation}

Furthermore, let $f'_*$ be the minimum value attained in the
feasible region (that is, at $p$), we know 
\begin{equation}
\label{eqn:nesterov_step4}
 f'_* = \Sigma_{\alpha \in A}{f_{\alpha} p^{\alpha}} 
\end{equation}


By assumption, $\tau_j$s are constant hence there exists a constant $\beta$ such that $\tau^\alpha \geq \beta$ for all $\alpha$. 
Combining~\eqref{eqn:nesterov_step1},
~\eqref{eqn:nesterov_step2},
~\eqref{eqn:nesterov_step3},
~\eqref{eqn:nesterov_step4},
we conclude that 
\[f'^*_k \leq \beta \frac{k!}{(k-d)!k^{d}} f'_* \]

Therefore, we get a constant factor approximation as 
\[  f'^*_k - f'_* \leq \left(1 - \beta \frac{k!}{(k-d)!k^{d}}  \right) (-f'_*)  \]

Note that when scaling down random walk from $x(p)$ to
$x'(p)$, we essentially move grid points beneath the simplex
(that is, into the region between the simplex, and axis planes).
Technically, these points are not in feasible region. However,
we notice that in IMLS', all coefficients in the objective function 
are non-negative. Thus,
in the minimization problem corresponding to IMLS', all coefficients
are non-positive. We can then essentially ``move up'' grid
points beneath the simplex onto the simplex plane by increasing
values. Since all coefficients are non-positive, this will not hurt
our objective function, we will get at least as good a value as $f^*_k$.

Thus, we will be able to find feasible points in the simplex using the
minimum value among all grid points, and obtain a constant
approximation factor.
\end{proof}
}

\condcomment{\boolean{fullpaper}}
{
\section{Proof of Theorem \ref{1}} 
\label{sec:idls_tricriteria}

\begin{proof}
We first obtain the following formulation IDLS$^\textrm{m}$
equivalent to IDLS.
Again we use $\hat{y}_i = 1 - y_i$ be the complement of $y_i$,
which indicates whether query $Q_i$ is un-selected.

\begin{align}
 \hspace{-1cm} {\text{(IDLS$^\textrm{m}$)}} \qquad{} \min \sum_{Q_i \in \mathcal{Q}} & { n_i w_i \hat{y}_i } \notag{}  \\
 \mbox{s.t.} \sum_{E_j \in \Sigma} & {\lambda_j m_j x_j} \leq M  \notag{} \\
 \sum_{Q_i \in \mathcal{Q}} &  { n_i c_i (1-\hat{y}_i) \leq C} \label{eqn:idls*_cpu} \\
 & \hat{y}_i \geq 1 - x_j, \forall {E_j \in Q_i}  \notag{} \\
 & \hat{y}_i, x_j \in \{0, 1\}  \notag{}
\end{align}

Since we have shown in Theorem~\ref{thm:imls_bicriteria}
that given a parameter $\frac{1}{\tau}$, solving the 
LP-relaxed version of IDLS$^\textrm{m}$ and then make selection
decision based on the rounding of fractional solutions
$x^*_j$ and $\hat{y}^*_i$ ensures that
memory consumption does not exceed $\frac{M}{1-\tau}$
and utility does not exceed $\frac{l^*}{1-\tau} $, what left
to be shown is that CPU consumption does not exceed 
$\frac{C}{1-\tau}$.

We again divide queries $\mathcal{Q}$ into the
promising set of queries
$\mathcal{Q}^a = \{ Q_i \in \mathcal{Q} | \hat{y}^*_i \leq {\tau} \}$
and the unpromising set
$\mathcal{Q}^r = \{ Q_i \in \mathcal{Q} | \hat{y}^*_i > {\tau} \}$. 

For the set of selected
queries, $\mathcal{Q}^a$, from~\eqref{eqn:idls*_cpu}
we know
\begin{equation}
\label{eqn:idls*_cpu_1}
 \sum_{Q_i \in \mathcal{Q}^a}{ n_i c_i (1-\hat{y}_i) }
 \leq \sum_{Q_i \in \mathcal{Q}}{ n_i c_i (1-\hat{y}_i)}
 \leq C
\end{equation}

Since $\forall Q_i \in \mathcal{Q}^a$, $\hat{y}_i \leq \tau$,
\begin{equation}
\notag{}
 \sum_{Q_i \in \mathcal{Q}^a}{ n_i c_i (1-\tau) } \leq
 \sum_{Q_i \in \mathcal{Q}^a}{ n_i c_i (1-\hat{y}_i) }
\end{equation}

Combining with~\eqref{eqn:idls*_cpu_1} we get
\begin{equation}
\notag{}
 \sum_{Q_i \in \mathcal{Q}^a}{ n_i c_i (1-\tau) } \leq C
\end{equation}

Because $\sum_{Q_i \in \mathcal{Q}^a}{ n_i c_i }$ is
the total CPU consumption, we can obtain that
\begin{equation}
\notag{}
 \sum_{Q_i \in \mathcal{Q}^a}{ n_i c_i  } \leq \frac{C}{1-\tau}
\end{equation}
This bounds the CPU consumption and completes our proof.
\end{proof}
}

\condcomment{\boolean{fullpaper}}
{
\section{Proof of Theorem~\ref{2}}
\label{sec:idls_approximable}

\begin{proof}
We define the following problem variant IDLS$^p$.

\begin{align}
\hspace{-1cm} \text{(IDLS$^p$)} \qquad{} \max & \sum_{Q_i \in \mathcal{Q}}  { n_i w_i y_i } \notag \\ 
 \mbox{s.t.} 
& \sum_{Q_i}{y_i (\sum_{E_j \in Q_i} {\lambda_j m_j})} \leq p M \label{eqn:idls^p_knapsack} \\
& \sum_{Q_i \in \mathcal{Q}}  { n_i c_i y_i } \leq C \notag \\
 ~~ & y_i \leq \prod_{E_j \in Q_i}{x_j} \notag \\
 & x_j, y_i \in \{0, 1\}  \notag
\end{align}

Because we know each event is shared by at most $p$ number
of queries, so for any solution $y_i, x_j$ to IDLS we have
\begin{equation}
\sum_{Q_i}{y_i (\sum_{E_j \in Q_i} {\lambda_j m_j})} \leq p \sum_{E_j \in \Sigma} {\lambda_j m_j x_j} = p M
\end{equation}
This implies that a solution feasible to IDLS must 
be feasible to IDLS$^p$, because from ILDS to
IDLS$^p$ only~\eqref{eqn:idls^p_knapsack} is modified
and we have ensured that any solution $y_i, x_j$ to IDLS
must also satisfy~\eqref{eqn:idls^p_knapsack}.
Let $OPT$ and $OPT^p$ be the optimal value 
of IDLS and IDLS$^p$, we can guarantee
\begin{equation}
\label{eqn:quality_opt}
OPT^p \geq OPT
\end{equation}

Further introduce the problem IDLS$^{ns}$

\begin{align}
\hspace{-1cm} \text{(IDLS$^{ns}$)} \qquad{} \max & \sum_{Q_i \in \mathcal{Q}}  { n_i w_i y_i } \notag \\ 
 \mbox{s.t.} 
& \sum_{Q_i}{y_i (\sum_{E_j \in Q_i} {\lambda_j m_j})} \leq M \label{eqn:idls^ns_knapsack} \\
& \sum_{Q_i \in \mathcal{Q}}  { n_i c_i y_i } \leq C \notag \\
 ~~ & y_i \leq \prod_{E_j \in Q_i}{x_j} \notag \\
 & x_j, y_i \in \{0, 1\}  \notag
\end{align}

We show that there exists a solution to IDLS$^{ns}$
that has a value of $\frac{1-\frac{1}{f}}{p} OPT^p$.
We construct this as follows. First initialize the solution 
to IDLS$^{ns}$ by setting $y_i^{ns} = 0, \forall Q_i$. 
Given the optimal solution $y_i^{p*}, x_j^{p*}$
to IDLS$^p$, we order non-zero $y_i^{p*}$s by 
$\frac{n_i w_i}{\sum_{E_j \in Q_i} {\lambda_j m_j}}$
descendingly. 
We can then iterate through non-zero $y_i^{p*}$s using that
order, and set the corresponding $y_i^{ns}$ to 1, 
and stop when picking the next item violates
the budget requirement, so that $y_i^{ns}$
respects constraint~\eqref{eqn:idls^ns_knapsack}.
\[
\sum_{Q_i}{y_i^{ns} (\sum_{E_j \in Q_i} {\lambda_j m_j})} \leq M
\]
It apparently also satisfies other constraints of IDLS$^{ns}$
because they are not changed from  IDLS$^{p}$ to  IDLS$^{ns}$.

Let this solution to  IDLS$^{ns}$ has value $v^{ns}$. We
know
\[
v^{ns} \frac{d}{1 - \frac{1}{f}} \geq OPT^{p}
\]
This is the case because first, given that 
$f = \frac{M}{\max_{Q_i}{\sum_{E_j \in Q_i}{m_j \lambda_j}}} $, using our construction, the slack of 
$M - \sum_{Q_i}{y_i^{ns} (\sum_{E_j \in Q_i} {\lambda_j m_j})}$
is at most $\frac{M}{f}$, or we have used 
at least $M - \frac{M}{f}$ memory. In addition, since
we pick $y_i^{ns}$ by their value to weight
ratio $\frac{n_i w_i}{\sum_{E_j \in Q_i} {\lambda_j m_j}}$,
increasing memory budget from $y_i^{ns}$ to $y_i^{p}$
by a factor of $p$ increases value by no more than 
a factor of $p (\frac{1}{1 - \frac{1}{f}})$.

Given $OPT^{ns } \geq v^{ns }$, we have
\[
OPT^{ns } \geq \frac{1 - \frac{1}{f}}{p} OPT^{p}
\]
Combining this with~\eqref{eqn:quality_opt} we know
\[
OPT^{ns} \geq \frac{1 - \frac{1}{f}}{p} OPT
\]

Notice that ILDS$^{ns}$ is the well-studied 
multi-dimensional knapsack problem
with dimensionality equals 2. Such problems can be solved in 
pseduo-polynomial-time
to obtain the optimal value $OPT^{ns}$.
Since the solution so produced is guaranteed to be within
$\frac{1 - \frac{1}{f}}{p} OPT$, our proof is thus complete.
\end{proof}
}

\end{document}